\documentclass[twocolumn]{aastex631}
\usepackage{newtxtext,newtxmath}
\usepackage[OT2,T1]{fontenc}
\usepackage{CJKutf8}

\DeclareRobustCommand{\VAN}[3]{#2}
\let\VANthebibliography\thebibliography
\def\thebibliography{\DeclareRobustCommand{\VAN}[3]{##3}\VANthebibliography}

\usepackage{graphicx}	
\usepackage{amsmath}	
\usepackage{color}
\usepackage{booktabs}
\usepackage{tcolorbox}
\usepackage{multirow}

\DeclareSymbolFont{cyrletters}{OT2}{wncyr}{m}{n}
\DeclareMathSymbol{\Sha}{\mathalpha}{cyrletters}{"58}

\newcommand{\newedit}[1]{{#1}}

\newcommand{\Imcom}{{\sc Imcom}}
\newcommand{\Pyimcom}{{\sc PyImcom}}
\newcommand{\Treecor}{{\sc TreeCorr}}
\newcommand{\Metacalibration}{{\sc Metacalibration}}

\shorttitle{Noise biasing for \textit{Roman} WL}
\shortauthors{K. Laliotis et al.}

\begin{document}

\title[Noise biasing for Roman WL]{Analysis of biasing from noise from the {\textit{\textbf{Nancy Grace Roman Space Telescope}}}: implications for weak lensing}

\correspondingauthor{Katherine Laliotis}
\email{laliotis.2@osu.edu}

\author{Katherine Laliotis}
\affiliation{Department of Physics,
The Ohio State University,
191 West Woodruff Avenue,
Columbus, OH 43210, USA}
\affiliation{Center for Cosmology and Astroparticle Physics,
The Ohio State University,
191 West Woodruff Avenue,
Columbus, OH 43210, USA}

\author{Emily Macbeth}
\affiliation{Department of Physics,
The Ohio State University,
191 West Woodruff Avenue,
Columbus, OH 43210, USA}
\affiliation{Center for Cosmology and Astroparticle Physics,
The Ohio State University,
191 West Woodruff Avenue,
Columbus, OH 43210, USA}
\affiliation{Department of Astronomy,
The Ohio State University,
140 West 18th Avenue,
Columbus, OH 43210, USA}

\author{Christopher M. Hirata}
\affiliation{Department of Physics,
The Ohio State University,
191 West Woodruff Avenue,
Columbus, OH 43210, USA}
\affiliation{Center for Cosmology and Astroparticle Physics,
The Ohio State University,
191 West Woodruff Avenue,
Columbus, OH 43210, USA}
\affiliation{Department of Astronomy,
The Ohio State University,
140 West 18th Avenue,
Columbus, OH 43210, USA}

\author{Kaili Cao (\begin{CJK*}{UTF8}{gbsn}曹开力\end{CJK*}$\!\!$)}
\affiliation{Department of Physics,
The Ohio State University,
191 West Woodruff Avenue,
Columbus, OH 43210, USA}
\affiliation{Center for Cosmology and Astroparticle Physics,
The Ohio State University,
191 West Woodruff Avenue,
Columbus, OH 43210, USA}

\author{Masaya Yamamoto}
\affiliation{Department of Physics, Duke University, Box 90305, Durham, NC 27708, USA}
\affiliation{Department of Astrophysical Sciences, Princeton University, Princeton, NJ 08544, USA}

\author{Michael Troxel}
\affiliation{Department of Physics, Duke University, Box 90305, Durham, NC 27708, USA}



\begin{abstract}
The {\slshape Nancy Grace Roman Space Telescope}, set to launch in 2026, will bring unprecedented precision to measurements of weak gravitational lensing. Because weak lensing is an inherently small signal, it is imperative to minimize systematic errors in measurements as completely as possible; this will ensure that the lensing measurements can be used to their full potential when extracting cosmological information. In this paper, we use laboratory tests of the \textit{Roman} detectors, simulations of the \textit{Roman} High Latitude Survey observations, and the proposed \textit{Roman} image combination pipeline to investigate the magnitude of detector read noise biasing on weak lensing measurements from Roman. First, we combine lab-measured detector noise fields with simulated observations and propagate these images through the \textit{Roman} image combination pipeline, \Imcom. We characterize the specific signatures of the noise fields in the resultant images and find that noise contributes to the combined images most strongly at scales relevant to physical characteristics of the detector including PSF shape, chip boundaries, and roll angles. We then measure shapes of simulated stars and galaxies and determine the magnitude of noise-induced shear bias on these measurements. We find that star shape correlations satisfy the system noise requirements as defined by the \textit{Roman} Science Requirements Document.
However, for galaxies fainter than $m_{\rm AB}\simeq24$, correction for noise correlations will be needed in order to ensure confidence in shape measurements in any observation band.
\end{abstract}

\section{Introduction}

Cosmology surveys of the next decade aim to solve large-scale questions about the nature of our universe. Gaining an understanding of cosmic acceleration, the nature of dark energy, and the behavior of gravity on large scales are just a few of these broad goals. Of the many proposed methods for probing these topics, one of the most promising for the immediate future is weak gravitational lensing \citep[WL;][]{Hoekstra2002, Weinberg2013, Mandelbaum2018}. 

In gravitational lensing, we observe that the shapes of distant galaxies and galaxy clusters are distorted; the light we receive from these distant objects is bent by large gravitating bodies as it travels towards us along our line of sight. For weak lensing, the observed gravitational lensing signal is very small --- on the order of $1\%$. Because of this, rather than looking at how much each object is sheared, we look for correlations in the measured shapes of multiple galaxies in the image.  This allows direct inference of information about the mass distribution along the line of sight, without dependence on what form that mass is in.
In combination with the Planck survey results \citep{PlanckCollab2020} and other recent and upcoming Cosmic Microwave Background (CMB) observations, WL will help us to test the accuracy of our current $\Lambda$CDM cosmological model against proposed alternatives (e.g., quintessence, modified General Relativity) \citep{Lemos2021}.

The very small magnitude of the weak lensing signal implies that in order to measure the cosmological parameters, a survey requires both a large sample size (hence the use of wide-field telescopes) and tight control over systematic errors arising from instrument signatures, image reduction and analysis, and theoretical modeling of the cosmological signal. Current WL surveys such as the Kilo Square Degree survey (KiDS; \citealt{deJong2013}), the Dark Energy Survey (DES; \citealt{Abbott2022}), and the Hyper Suprime Cam (HSC; \citealt{Aihara2018}) have reached an accuracy of a few percent on the overall amplitude parameter, $S_8$ \citep{Amon2023}. 

With the next generation of weak lensing surveys, including the \textit{Euclid} wide field visible and near-infrared (NIR) surveys \citep{Laureijs2011, Scaramella2022, Euclid2024}, the Vera C. Rubin Observatory (\textit{Rubin}) \newedit{Legacy Survey of Space and Time (LSST)} \citep{Ivezic2019}, and the \textit{Nancy Grace Roman Space Telescope} (\textit{Roman}) High-Latitude Imaging Survey (HLIS) \citep{Spergel2015, Akeson2019}, this accuracy will improve significantly. This will be accomplished through improvements in survey field sizes and depths, increasing statistical constraints on WL signals, and through refined systematic error mitigation, increasing the precision of the signals themselves.

\textit{Roman} and \textit{Euclid} are both space-based surveys, giving them higher angular resolution than the ground-based \textit{Rubin}. However, the high angular resolution results in several complications. The point-spread function (PSF) of a circular-aperture telescope is determined by $\lambda/D$, where $D$ is the telescope diameter and $\lambda$ the observed wavelength.\footnote{Technically, the highest spatial frequency Fourier mode that is preserved by the optics can be no larger than 1 cycle per $\lambda/D$.} If the pixel grid spacing is $>\lambda/2D$, the detected signal will be \textit{undersampled} by the Nyquist criterion. Because of resource and technical limitations (one wants to survey the sky as fast as possible, but is limited by practical considerations of pixel count and overheads), both \textit{Roman} and \textit{Euclid} are undersampled \newedit{in single exposures}.
 
There are several already existing image processing tools that can be applied to calibrate WL shape measurements even without prior knowledge of the ``true'' distribution of intrinsic galaxy shapes, including \Metacalibration\, \citep{Huff2017}, Bayesian Fourier Domain \citep[e.g.][]{Bernstein2014, Bernstein2016}, and analytical methods \citep{Li2023}. However, all of these methods depend on the use of fully sampled images, which can be losslessly resampled, translated, and sheared. \newedit{\citet{Kannawadi2021} showed that running \Metacalibration\, on undersampled images results in significant biases caused by aliasing.}

In order to meet the exacting requirements for shear calibration error defined by the \textit{Roman} Science Requirements Document (SRD) \footnote{Originally derived during the Concept and Technology Development phase, WL requirements are found in \citet{Dore2018} or \url{https://asd.gsfc.nasa.gov/romancaa/docs2/RST-SYS-REQ-0020C_SRD.docx}, the latter being an updated version of the former.}, we need a method of reconstructing images that are well-sampled to be processed through shear measurement calibration.
This is accomplished through image combination: multiple undersampled dithered exposures of each field of view are combined together into one image using linear algebra-based techniques. 

\Imcom\,\ \citep{Rowe2011} is an example of one such linear algebra image combination algorithm that generalizes previous work \citep{Lauer1999a, Lauer1999b} to allow for rotations and missing samples. It is designed to find an optimal matrix that maps the undersampled input pixel values to a grid of oversampled output pixels, all while minimizing both the noise in the output image and the deviation of the true output PSF from some desired ``target'' PSF. The initial application of \Imcom\, to \textit{Roman} simulations \citep[hereafter T23]{Troxel2023} was presented in \citet[hereafter H24]{Hirata2023}, and the WL systematics in \Imcom\,\ outputs were characterized in \citet[hereafter Y24]{Yamamoto2023}. In this work we focus on results of a new set of simulations, OpenUniverse2024 (OU24)\footnote{Data preview now available at: \url{https://irsa.ipac.caltech.edu/data/theory/openuniverse2024/overview.html}}, processed with
\Pyimcom\,: a new implementation of the {\sc Imcom} image coaddition algorithm \citep{2024arXiv241005442C}. We choose to focus on the OU24+\Pyimcom\, data set because it is the most similar to the current plan for \textit{Roman} images and image processing. However, in the interest of completeness, we include a discussion of results from the T23+\Imcom\, combination as well in Appendix \ref{app:T23}.

The two sets of analysis mentioned above also use two different sets of detector read noise data, described in \S\ref{sec:labdata}. We combine real detector noise data from the Detector Characterization lab (see \S\ref{sec:labdata}) with simulated \textit{Roman} observations to determine a) how noise fields from the real hardware propagate through image coaddition to shape measurement and b) expected levels of noise biases in the shape correlation function. Ultimately, we determine how and by how much noise correlations will need to be corrected if we are to use \Pyimcom\, to process the real \textit{Roman} data.

\begin{figure*}
    \centering
    \includegraphics[width=0.8\linewidth]{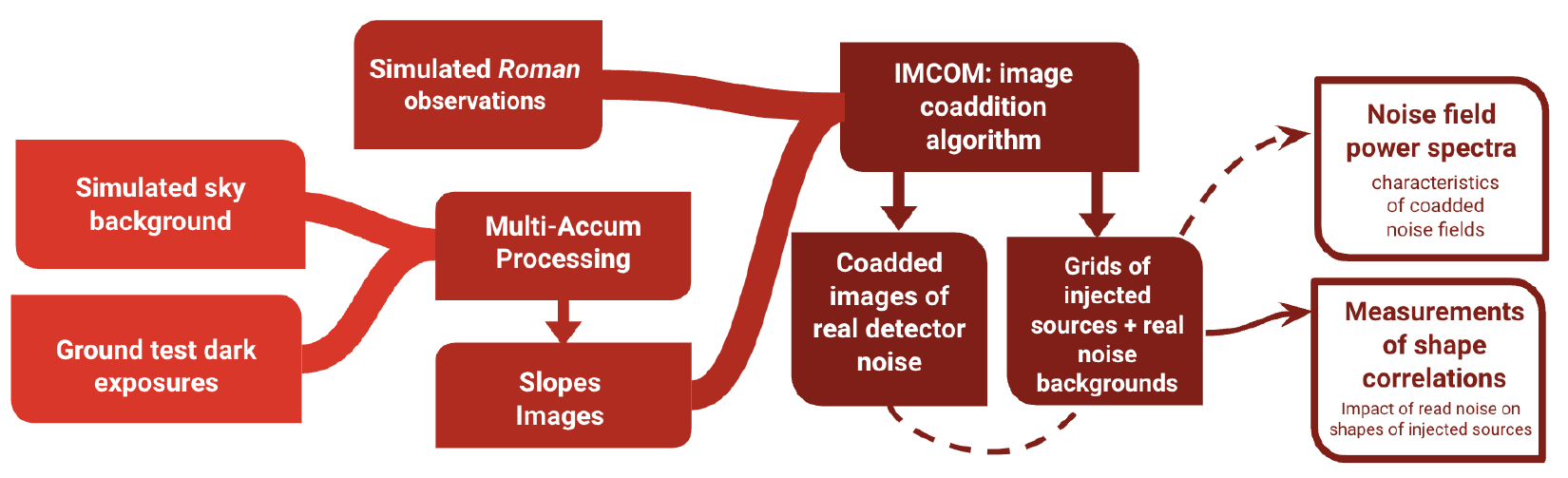}
    \caption{A flowchart illustrating the steps taken in this work.}
    \label{fig:flowchart}
\end{figure*}

This paper is organized as follows. In Section~\ref{sec:labdata}, we describe the two sets of laboratory test data of the flight detector system, along with preliminary processing steps taken to simulate how the frames will be sampled and processed on \textit{Roman} before image combination back on Earth. Section~\ref{sec:simdata} contains a brief description of the image simulations and outlines the method of ``injecting'' stars and galaxies whose shapes we use to constrain noise biasing.
In Section~\ref{sec:noise_corr}, we present correlations of detector noise fields via both 1- and 2-dimensional power spectra, and in Section~\ref{sec:noise_bias} we estimate the additive bias from read noise on shape measurements of simulated sources from OU24-sims+\Pyimcom\,. Finally, in Section~\ref{sec:discussion}, we discuss the results of this analysis and suggest next steps for future works. Figure \ref{fig:flowchart} illustrates the workflow of this project.

\section{Laboratory Data}
\label{sec:labdata}

\subsection{Overview}
\label{subsec:labdata}

The data for this study comes from ground testing of the \textit{Roman} detectors. The \textit{Roman} Wide Field Instrument (WFI) contains 18 Sensor Chip Assembly (SCA) detectors. Each detector contains a light-absorbing Hg$_{1-x}$Cd$_x$Te semiconductor, connected to a silicon readout integrated circuit (ROIC) via indium interconnects \citep{Mosby2020}. The $4096\times 4096$ pixels are divided into 32 readout channels, with 128 columns of pixels per channel. The outer four rows and columns of each SCA are reference pixels (these do not sense light), making the ``active region'' of the detectors a $4088\times 4088$ grid. Each pixel on these infrared detector arrays is a photodiode, and as a pixel is exposed to light and charge builds up, the voltage on the side of the photodiode facing the ROIC increases. The ROIC allows each pixel to be addressed and the voltage to be sensed non-destructively (via the gate of a field effect transistor), so we can read out the signal $S$ (units: data numbers or DN) multiple times over the exposure. In addition to the 32 science readout channels, the SCA provides a reference output, which does not correspond to a physical pixel but has some common-mode noise with the science outputs. The ACADIA controller \citep{2018SPIE10709E..0TL} provides biases, clocking signals, etc.\ to the SCA via a flexible cable, and receives the analog voltages back from the SCA. The signal is digitized by 16-bit analog-to-digital converters (ADCs) in the ACADIA. Including the reference output, which acts as a 33rd channel and is digitized at the same time, the resulting frames make a $4096\times 4224$ array (of which a $4088\times 4088$ subregion are the active pixels).

Testing of the SCAs is carried out at several levels of integration. Each SCA delivered undergoes {\em acceptance testing} at the Detector Characterization Laboratory (DCL) at NASA Goddard Space Flight Center. Acceptance testing includes standard measurements (e.g., dark current, noise performance, flat field, linearity, cross-talk, persistence, quantum efficiency), and is carried out with a laboratory (Generation 3 Leach) controller. The SCAs that are most promising for the flight program then undergo {\em triplet testing}. In this stage SCAs are connected via flex cables to an ACADIA controller, the operational parameters are tuned to optimize the SCA+cable+ACADIA system, and additional characterization data are obtained. The third stage is the {\em focal plane system testing}, with the 18 flight SCAs and controllers installed in the focal plane system.

The initial pixel mask and performance parameters in \citet{Troxel2023} were derived from the acceptance tests. This paper uses dark and flat data from the triplet tests and the focal plane system tests (analyzed separately). The triplet test data were acquired during the period 2021 Jun 14 through 2021 Nov 06 for the 18 SCAs that were selected to fly in the focal plane. These tests were pre-processed and included in the T23-sims+\Imcom\, runs (analyzed in H24 and Y24), though the triplet test pre-processing was only finished in time for inclusion in the J129 and H158 bands. Results from the triplet tests are included in Appendix \ref{app:T23}.

The focal plane system test (FPT) data were acquired in 2023 Apr 28--29 during the final performance test before delivery of the focal plane for integration into the Wide Field Instrument. (This was actually the second of two focal plane tests; in between, environmental testing was carried out, and three of the SCAs were replaced with spares.) These tests were pre-processed and included in the OU24-sims+\Pyimcom\, runs, and their analysis constitutes the main results of this paper. We chose to focus on the FPT noise realisations as they are the most flight-like and therefore will give the most realistic expectations for noise biasing from read noise.

A more detailed description of the characterization and selection of flight detectors can be found in \cite{Mosby2024}. Data from the DCL is distributed via the Advanced Data Analytics PlaTform (ADAPT) at the NASA Center for Climate Simulation, and available for use to the \textit{Roman} Image Simulation team.

\subsection{Preliminary Processing}
\label{subsec:preprocess}

The first levels of processing applied to the data before analysis consist of: addition of sky background (\S\ref{subsubsec:skybckgnd}), Multi-Accum processing (\S\ref{subsubsec:multiaccum}), weighted Least-Squares fitting (\S\ref{subsubsec:LSF}), and reference pixel corrections (\S\ref{subsubsec:refpixel}).

\subsubsection{Sky Background}
\label{subsubsec:skybckgnd}

In addition to the read noise inherent to the SCAs, which we study by looking at the dark frames, each readout will contain some background noise from the sky, which will also accumulate with each frame because of the ROIC nondestructive reading. This sky background is another potential source of biasing in observations.

Photons arriving randomly to the detector at some rate fall in a Poisson distribution, centered around the mean expected signal for that pixel. The expected sky\newedit{+thermal} background observing in the H band at high ecliptic latitude is 0.38 e pix$^{-1}$ s$^{-1}$ \newedit{(as determined by the Exposure Time Calculator, \citealt{2012arXiv1204.5151H} at $\pm75^\circ$ ecliptic latitude and at quadrature)}. For frames exposed over 3.08 s (frame time in flight mode for WFI SCAs), this gives a mean sky background of 1.17 e pix$^{-1}$. 

We simulate this sky background as a random Poisson-distributed noise map centered around the mean signal. After each frame $\alpha = \{0,1,...49\}$ is read out, a new realization of the sky background noise map $P_\alpha$ is added to frames $\alpha$ through the final frame $N_{\alpha}=50$, i.e.,
\begin{equation}
S_\alpha = S_\alpha + \sum^\alpha_{\beta=0} P_\beta.
\end{equation}

We note that there may be some additional nonlinear effects from sky noise which are not modeled in this work; we leave their inclusion to a future noise analysis work. 

\subsubsection{Multi-Accum Processing}\label{subsubsec:multiaccum}

The \textit{Roman} WFI SCAs are read non-destructively, so with each readout, the pixels are not reset and instead continue to accumulate charge. A ``frame'' of data (reading the whole SCA) takes 3.08 seconds, so some multiple of this constitutes an exposure. At the time this project was started, the plan was for an exposure to consist of 50 frames, so that will be assumed in this section. We note, however, that this number is being revisited during the survey definition process and may not be final. 

In an ideal world, we could collect data from and use all 50 frames per exposure. For Roman's $\sim300$ Mpix focal plane, sampling 16 bits per pixel with $\sim 3$ s exposures gives a raw data rate of $\sim1.6$ Gbps. However, we will be limited by the real-world constraint of the speed at which data can be transmitted down to Earth from L2, where \textit{Roman} will be operating. The \textit{Roman} downlink rate (as of Jan. 2023) is 275 Mbps\footnote{\url{https://github.com/spacetelescope/roman-technical-information/tree/v1.0/data/Observatory/MissionandObservatoryTechnicalOverview}}.

This fact necessitates the development of a ``Multi-Accum'' processing step \citep[\S7.7]{Dressel2022}: combining together multiple frames in some optimal way and creating ``averaged'' frames containing as much information as possible. This process reduces the amount of data needing to be downlinked, and also helps to eliminate biasing from read noise. In this work, we combine \textit{Roman} test images into 6 Multi-Accum frames per exposure.

The choices of frames are informed by several criteria. The beginning and end of the up-the-ramp sequence give the highest signal-to-noise ratio on bright but unsaturated sources, and the beginning is important for sources that will saturate.
We also want to include some middle-time frames, to have a central point in the up-the-ramp sequence to be able to flag cosmic rays.
For this project, we choose the set of frames that results in optimal accuracy for shape measurement; we note that this choice will not necessarily be the same in the final \textit{Roman} image processing pipeline.

A first-approximation optimization of frame choice was done by comparing a shear measurement on several different choices of frames following these loose guidelines. We selected several different sets of raw frames to average together into five or six Multi-Accum frames. Next, we used the method of least squares to reduce the 5 or 6 Multi-Accum frames into a single slope image (see Section \ref{subsubsec:LSF}). 

To select which frames that will optimize measurements of galaxy shapes, we want to find the frames that will minimize variance in shear, $\Delta \mathbf{\eta}$. We follow \cite{Bernstein2002}, calculating the change in the quadrupole term of measured shape $\delta M$ to find the ellipticity components
\begin{equation}
    \Delta e_{1}=\frac{4}{f\sigma^{2}}\int (\Delta x^{2}-\Delta y^{2}) e^{-\frac{\Delta x^{2}+\Delta y^{2}}{2\sigma^{2}}}  I(x-\Delta x, y-\Delta y)d\Delta xd\Delta y
\end{equation}
and
\begin{equation}
    \Delta e_{2}=\frac{4}{f\sigma^{2}}\int (2\Delta x \Delta y) e^{-\frac{\Delta x^{2}+\Delta y^{2}}{2\sigma^{2}}}  I(x-\Delta x, y-\Delta y)d\Delta xd\Delta y
\end{equation}
across the full slope image. 
The full derivation of these equations can be found in Appendix \ref{app:e_derivation}. These are evaluated via convolution with the input frames. We compute the means and variances of these ellipticity measurements on several candidate slope images, which are made up of different frame selection choices. We select the set of frames whose resulting slope image minimized the variances of $\Delta e_1$ and $\Delta e_2$.

Input frames for each Multi-Accum output frame were combined together using simple averaging (as will be implemented on board \textit{Roman}). For each exposure, we first read in the full set of frames $S_{\alpha}$, ($ \alpha=0,1...49$). Then, for each Multi-Accum (MA) output frame $\mathbb{S}_{\beta}$, ($\beta=0,1...5$), we sum up the values in the pixels and divide by the number of frames being included in the MA frame (N$_{\beta}$):
\begin{equation}
\mathbb{S}_{\beta\,} = \frac{1}{N_\beta} \sum_{\alpha=\alpha_\beta}^{\alpha_\beta+N_{\beta}-1} S_{\alpha\,}
\end{equation}

where $\alpha_\beta$ denotes the first frame in each MA frame.

We found minimal noise variance from the combination of frames: 
\\
\newedit{MA} Frame 0: (2) \\
\newedit{MA} Frame 1: (3,4) \\
\newedit{MA} Frame 2: (5,6,7,8) \\
\newedit{MA} Frame 3: (23,24,25,26)\\
\newedit{MA} Frame 4: (44,45,46,47)\\
\newedit{MA} Frame 5: (49),\\

where the frame indices, $\alpha_i$, used are indicated in parentheses, and adopt this choice for the remainder of the analysis in this paper. 

\begin{figure}[h]
    \centering
    \includegraphics[width=3in]{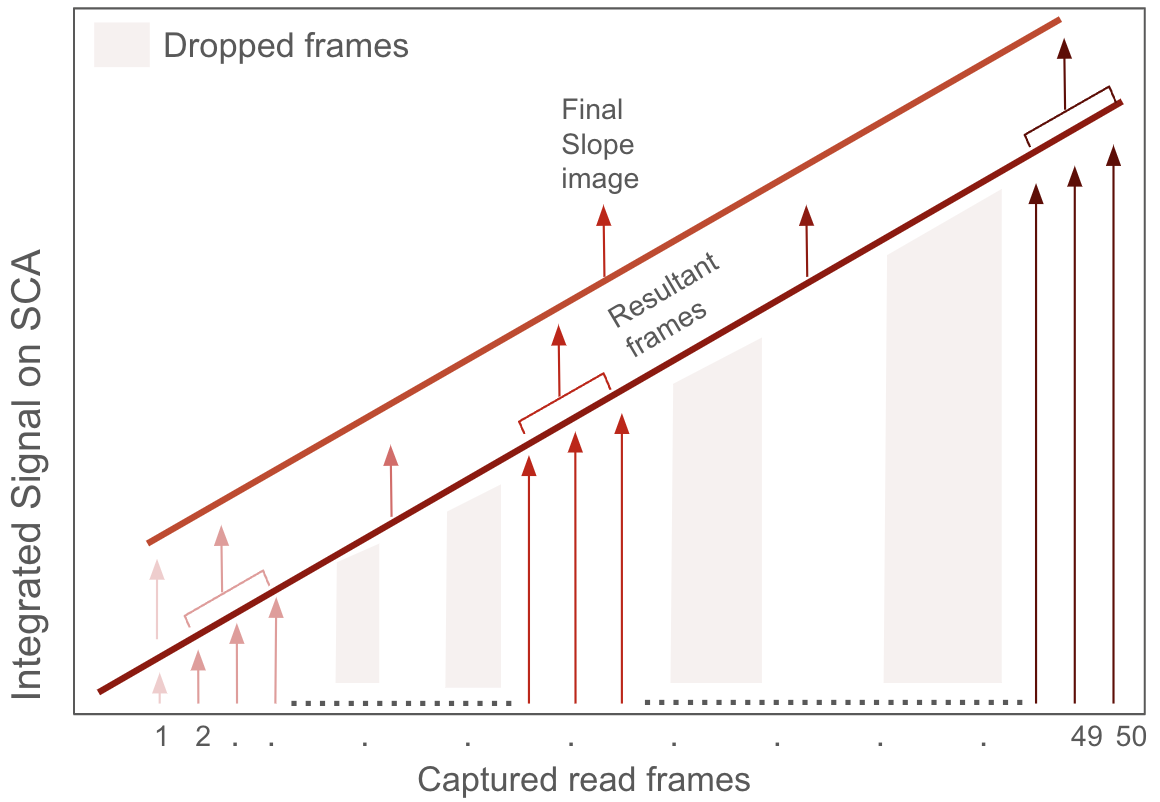}
    \caption{Illustration of the Multi-Accum and slope image process.}
    \label{fig:slopesimg}
\end{figure}

The output MA image $\mathbb{S}_{\alpha\,i,j}$ from this step is a 6-frame dark exposure. The MA and slope image construction is illustrated in Figure \ref{fig:slopesimg}.

\subsubsection{Least-Squares Fitting\label{subsubsec:LSF}}

We further compress the 6-frame MA image into a single frame by using a covariance-weighted least-squares fit. We define our output noise image as a linear combination of MA frames, $I^{\rm out} = \sum_{\alpha} W_{\alpha} \mathbb{S}_{\alpha\,ij}$, where $W_{\alpha}$ are weights to be determined by the least-squares fit.

First we flatten our 2D MA frames $\mathbb{S}_{\alpha\,,ij}, \text{ where } i,j=[0,4088-1]$ into 1D, $\mathbb{S}_{\alpha, i}, \text{ where } i=[0, 4088^2-1]$.
We then find the covariance matrix for each pixel between MA frames,
\begin{equation}
C_{\alpha\beta} = \frac{1}{N_{pix}-1} \sum_{i} (\mathbb{S}_{\alpha i} - \Bar{\mathbb{S}}_{\alpha i})(\mathbb{S}_{\beta i} - \Bar{\mathbb{S}}_{\beta i}),
\end{equation}
where $\Bar{\mathbb{S}}_{\alpha i}$ denotes the average signal across MA frames in pixel \textit{i}. 

We construct a $2\times6$ ``time'' basis vector $X_{A\alpha}$ containing values of different orders in $t_{\rm frame}$.
Capital roman letters denote orders of \textit{t}: $A=1$ and $B=t$, while lowercase Greek letters denote MA frame (eg., if $\mathbb{S}_3$ is an average of $S_5...S_8$, then $X_{13} = 6.5$) Using this vector basis we solve for the coefficients $b$:
\begin{equation}
    b = \Big[ XC^{-1}X^{T}\Big]^{-1}XC^{-1}
\end{equation}

 The vector b is $2\times4088^2$, the first dimension being the added constants and the second dimension being the slopes: $\mathbb{S}_i = b_{1,i}t + b_{0,i}$. The slope $b_{1,i}$ contains the information we want: each pixel's signal up-the-ramp over multiple frames. 

\begin{figure}[h]
    \centering
    \includegraphics[width=3.2in]{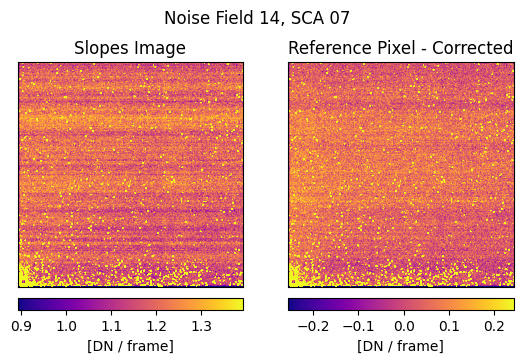}
    \caption{\label{fig:refcor} The bottom left $1000\times1000$ pixel region of a single noise realisation on SCA 7. Left: Uncorrected slope image; Right: Reference pixel-corrected slope image. \newedit{The colorbar shows DN/frame, with minimum and maximum values of each image's median $\pm0.25$. The noise stripes are lessened but not completely removed.}}
\end{figure}

Finally, we take the slope value $b_{1,i}$ for each pixel and construct a $4088\times4088$ image of the slopes of pixel signal with time. This ``slope image'' will be used to characterize the noise biases stemming from read and sky background noise. An example slope image is shown in the left panel of Figure \ref{fig:refcor}.

\subsubsection{Reference Pixel Corrections}\label{subsubsec:refpixel}

From the slope images, we can clearly see horizontal bands that extend through the SCA active region \textit{and} the reference pixels, indicating that they are caused by electronics rather than actual signal. In this section, we describe our method to first-order correction of this image striping using the reference pixels. We note that the \citet{Rauscher2022} method of reference pixel subtraction is more likely to be used for actual correction of \textit{Roman} image banding, and that in future work we will implement this improved method.

For each of the 32 readout channels (see Section \ref{sec:labdata}), we calculate the median pixel value in the reference pixel rows. Then a least-squares fit is computed, treating the medians endpoints. Using the fit function, we calculate and subtract out the expected electronic signal from each subsequent pixel row, moving down the active region row by row. We repeat this process for each active channel. 

Reference pixels are also useful to account for signal biasing due to the detectors operating in differential mode. As with the vertical channels, the detectors can have some electronic signal that grows linearly across the row, which we estimate by finding the median of the reference pixels this time on the leftmost and rightmost regions of the SCA and fitting a slope between the lines. We also calculate the median value of the active region pixels, and include that value as a y-intercept of our best fit line. As before, we subtract out the value of the best fit line at each pixel along the row.

An example of a slope image before and after including the above reference pixel correction scheme is shown in Figure \ref{fig:refcor}. The comparison shows that this method is effective at removing much of the vertical striping in the slope images, though not all. More sophisticated methods of reference pixel correction (e.g., \citet{Rauscher2022}) will be applied in future work.

\begin{figure}
    \centering
    \includegraphics[width=3.0in]{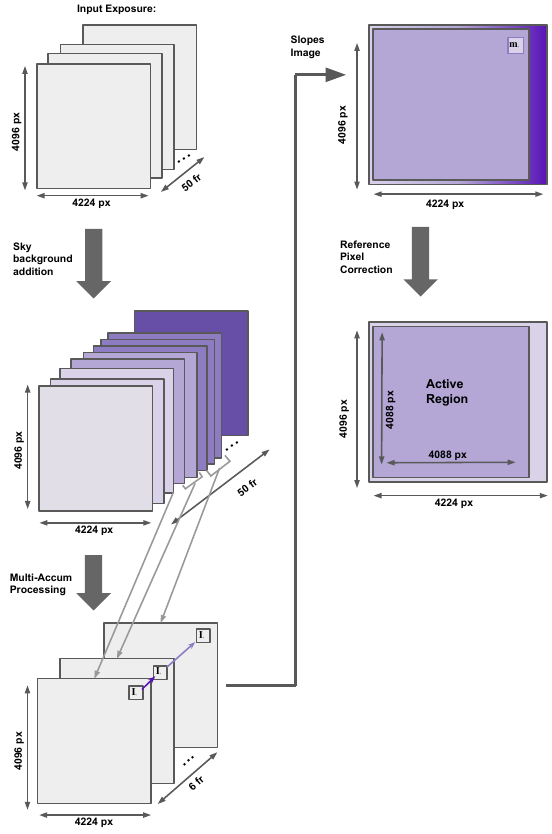}
    \caption{\label{fig:preprocess} This figure illustrates the pre-processing steps taken on the read noise data frames before their inclusion in this run of \Imcom\,. \newedit{Sky background addition results in additional signal (represented by purple shading) being added to each frame. Multi-accum processing condenses the 50 frames down to 6 frames. For a given pixel, the change in signal `I' over the MA frames is fit to a line with slope `m', which becomes the pixel value. Finally reference pixel corrections are applied to remove in-row stripes. In-depth discussion of each step can be found in Section }\ref{sec:labdata}.}
\end{figure}

\section{Simulated Data}
\label{sec:simdata}

Two sets of simulated data are used for this work. The triplet detector tests are combined with the joint Roman+Rubin simulations from \citet{Troxel2023} (T23 simulations). The FPT detector tests are combined with the new Roman+Rubin simulation OpenUniverse2024 (OU24; Troxel et al, in prep.). Once observations are simulated, they are combined with detector read noise data, filtered through a pixel mask, and combined through application of the \Pyimcom\, (\Imcom, for T23) image coaddition algorithm. For this work, we do not actually use the simulated survey images; however, due to the computational cost of running image combination, we processed our noise frames and injected star and galaxy grids in the same \Pyimcom\, run as the OU24 simulations. We include a brief discussion of the synthetic surveys below anyways, as future analyses from this group will make use of them.

\subsection{Synthetic Surveys}

In H24, \Imcom\, is run on the T23 simulations, which are based on the \textit{Roman} telescope properties and the current Reference Survey \citep{Spergel2015, Akeson2019}. The T23 simulations cover 20 deg$^2$ of the \textit{Roman} observation fields ($1\%$ of the total area from the \textit{Roman} Reference Survey), computed using the LSST DC2 simulated sky \citep{Korytov2019, LSSTDESC2021}. The area is integrated over the \textit{Roman} coverage region, including \textit{Roman} telescope and instrument designs from the Reference Survey \citep{Akeson2019}. Based on the Reference Survey, the \textit{Roman} simulated images are taken in the Y106, J129, H158, and F184 bands. Multiple dithered exposures of each field of view are generated in order to cover chip gaps and correct for other detector systematics. Additionally, the T23 simulations include detector physics specific to the \textit{Roman} SCAs, based on the detector characterization work from \cite{Mosby2020}. More detailed descriptions of the implementation of Reference Survey characteristics and detector physics into the simulated images are located in Appendices A and B of \cite{Troxel2023}.

H24 describes in detail the process of image combination, along with the specifics of the input and output simulation data, and shows the validity of using \Imcom\, as a method of image reconstruction for \textit{Roman}. Y24 goes a step further, taking these results and analyzing their impact on weak lensing measurements, including:

\begin{itemize}
    \item Measurements of noise correlations for simulated noise fields
    \item Moments and correlation functions analysis of sources in the coadded images and noise fields
    \item Construction and analysis of synthetic wide-band images 
\end{itemize}

In late 2023, a new suite of \textit{Roman}+\textit{LSST} simulations was developed by a collaboration of LSST-DESC and the \textit{Roman} High Latitude Imaging Survey (HLIS) \newedit{and High Latitude Time Domain Survey (HLTDS)} Project Infrastructure Teams (PITs) and run on the supercomputer at Argonne National Lab (publication in prep.). These OpenUniverse2024 (OU24) simulations' observation fields are completely overlapping for the two surveys. They also included two additional observing bands, K-band (K, F213) and a Wide band filter (W, F146) in the HLIS. Images in the wide filter were not processed by {\sc Imcom} due to difficulty with the chromatic PSF (Berlfein et al, in prep.) These simulations were then run through the new \Pyimcom\, image combination algorithm, described in detail in \citet{2024arXiv241005442C}. \Pyimcom\, largely follows the same outline as \Imcom\,, but is concisely organized into one convenient-for-use Python package. 

The OU24 simulations are both an expansion and an updated version of the T23 simulations. While the T23 simulations focused on \textit{Roman}'s HLIS, the OU24 simulations include overlapping Roman+Rubin transient time domain and wide and deep field surveys as well. Two additional observing filters (W146 and K213) are simulated, and the survey strategy is updated to be as useful as possible for future selection of the core survey. \newedit{The input pixel scale is kept the same as in T23, at $s_{\rm in}=0.11$ arcsec$^2$}. The updated \textit{Roman} wide and deep fields also contain updated versions of detector physics modeling, including a more sophisticated model of charge diffusion (explored in the forthcoming work by Macbeth et al, in prep.), and account for some known issues in the T23 simulations.  For these reasons, our analysis in Sections \ref{sec:noise_corr} and \ref{sec:noise_bias} focuses on the OU24-sims. Results of running this analysis on the T23+\Imcom\, combination are included in Appendix~\ref{app:T23}.

\begin{figure}[h]
    \centering
    \includegraphics[width=3.2in]{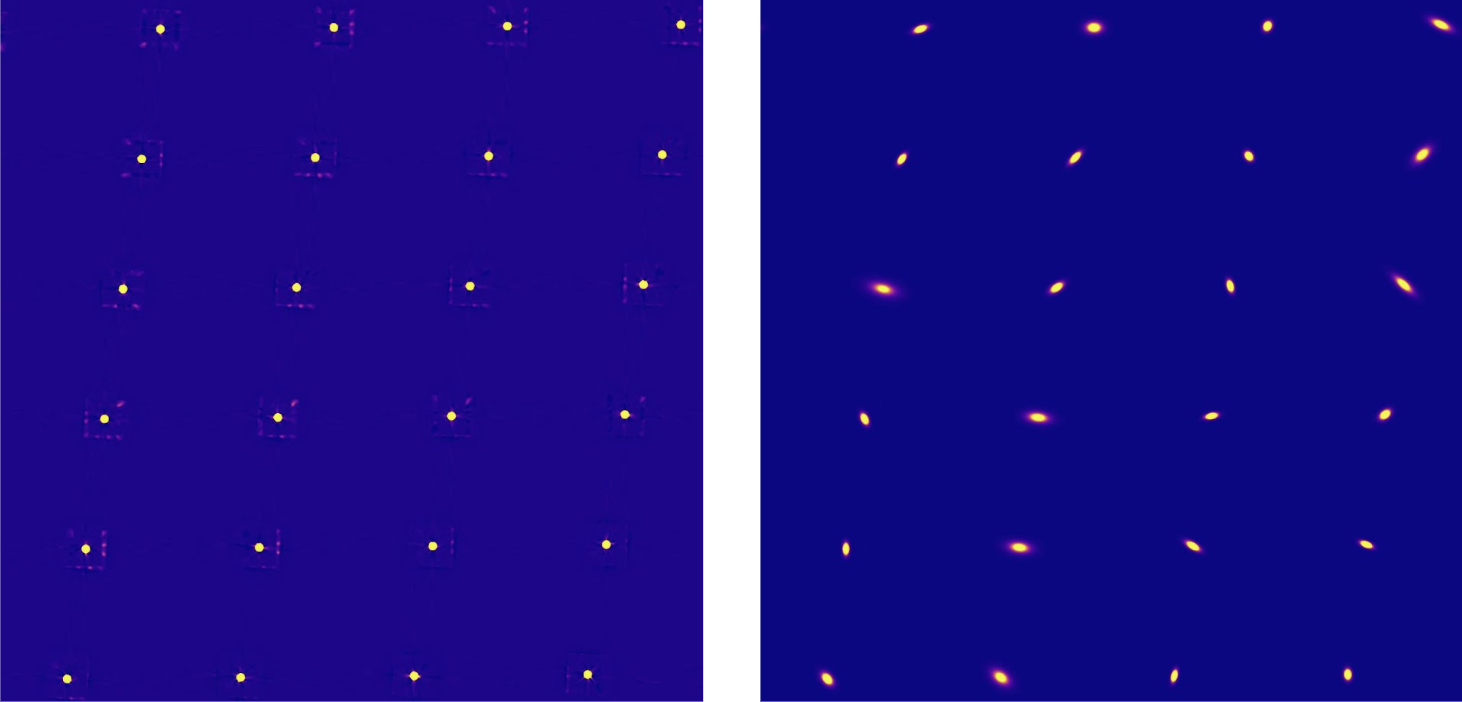}
    \caption{\label{fig:injectedobjects} Left (Right): Upper left quarter of {\sc GalSim} injected star (galaxy) grid in block (0,0) in the Y band.}
\end{figure}

Beyond the simulated surveys, we also use {\sc GalSim} \citep{Rowe2015} to draw a grid of simulated objects in each coaddition block. We refer to these object grids as ``injected'' stars and galaxies. All injected objects are drawn at HEALPix resolution of 14 (or NPixels $= 12 \times 4^{14}$). Stars are simply modeled as $\delta$ functions convolved with the {\sc GalSim} simulated Roman PSF. Injected galaxies are created with a Sersic profile with Sersic index $n=1$ and flux $F=1$, then sheared by a small amount in $g_1$ and $g_2$.  An example of each injected object grid is shown in Figure \ref{fig:injectedobjects}.

\subsection{Image Combination}

To include the impact of detector read noise, we next combine the reference pixel-corrected slope images of the noise fields (the product from Section \ref{sec:labdata}) with the simulated \textit{Roman} observations and injected source grids. The \Imcom\, algorithm \citep{Rowe2011} implements linear algebra-based combination methods, so it is valid to simply superimpose images together before inputting them into \Imcom\,. The observation+noise images are then filtered by a pixel mask, and finally combined through the \Pyimcom\ \citep{2024arXiv241005442C} implementation of \Imcom\,.

Pixel masks imposed on the images come from laboratory tests of the detectors. Pixels that are non-responsive, hot, high-noise, or have failed non-linearity or gain solutions are masked out, as are their nearest and diagonal-nearest neighbors. In addition, a simulated cosmic ray mask was generated as described in H24. This resulted in a total of 1.15\% of pixels being masked out. 
Imposition of the same pixel mask over every image means that each input layer can be processed simultaneously, so \Imcom\, computes the transformation matrix (from input image pixels to output combined image pixels) only once. 

The image coaddition process is designed to combine together multiple undersampled observations from \textit{Roman} into a fully-sampled image (mosaic). The input images may have different input pixel scales, pixel masks, orientations, distortions, and PSFs; construction of a coherent output image thus requires interpolation over masked regions, re-sampling pixels onto a common grid, homogenization of the output PSFs, and averaging of the various input images. The ``target'' output PSFs given to \Imcom\ are shown in Table~\ref{tab:target}. 

\begin{table}
    \centering
    \caption{\label{tab:target}The target PSFs used in this work. For each simulation, the target PSF type is shown, and then the parameters of that target PSF. The function for the obstructed Airy disc $*$ Gaussian with linear obstruction factor $\varepsilon$ is given in H24, Eq.~(5). The Full Width at Half Maximum (FWHM) is also shown.}
    \begin{tabular}{c|cccc|cc}
    \hline\hline
    Filter & \multicolumn4c{T23+\Imcom} & \multicolumn2c{OU24-sims+\Pyimcom} \\
     & \multicolumn4c{obstructed Airy + Gaussian} & \multicolumn2c{Gaussian} \\
     & $\lambda/D$ & $\varepsilon$ & $\sigma_{\rm Gauss}$ & $\theta_{\rm FWHM}$ & $\sigma_{\rm Gauss}$ & $\theta_{\rm FWHM}$ \\
     & \![arcsec]\! & & \![arcsec]\! & \![arcsec]\! & \![arcsec]\! & \![arcsec]\! \\ \hline
     Y106    & \multicolumn4c{not run} & 0.093 & 0.220\\
     J129    & 0.112 & \!0.31\! & 0.082 & 0.230 & 0.098 & 0.231\\
     H158    & 0.138 & \!0.31\! & 0.070 & 0.210 & 0.103 & 0.242\\
     F184    & \multicolumn4c{not run} & 0.107 & 0.253 \\
     K213    & \multicolumn4c{not run} & 0.112 & 0.264 \\
    \hline\hline
    \end{tabular}
\end{table}

\Pyimcom\,, an optimized version of the coaddition algorithm from H24, is now available for use\footnote{https://github.com/kailicao/pyimcom}, and is described in detail in \citet{2024arXiv241005442C}.  Because of the faster runtime and overall improvements, the  focal plane test noise frames + OU24-Simulations are processed using \Pyimcom\,. Improvements include changes to some linear algebra techniques which improve the runtime of the coaddition process, and development of a more object-oriented framework. In particular, \citet{2024arXiv241005442C} explore the use of different linear algebra kernels during the coaddition process. This work uses the Cholesky decomposition kernel, as described in that paper. \citet{2024arXiv241005442C} confirm the accuracy of \Pyimcom\, by running it on the same simulations as H24, and show that essentially the same results are recovered. 

Additionally, informed by the H24 and Y24 analysis of T23+\Imcom\,, several parameter choices were changed in the implementation of \Pyimcom\,. One of those changes was the target output PSF, which in  H24 was specified as an Airy disk convolved with a Gaussian but in this work is simply a Gaussian. \Pyimcom\, also adopted a new strategy for determination of the Lagrange multiplier $\kappa_\alpha$, setting it to a constant value to improve boundary effects and coaddition efficiency.

Due to the computationally intensive nature of coaddition, the full $1\times 1$ degree mosaic images are subdivided into $36\times36$ blocks of $1.667 \times 1.667$ arcmin each, which are then further divided into a grid of $80 \times 80$ square ``postage stamps'' of $32 \times 32$ output pixels ($s_{\rm out} = 0.0390625$ arcsec). Postage stamps are coadded one by one and then added into the output mosaic. More detailed descriptions of the \Imcom\, and \Pyimcom\, algorithms, including reasoning for the change of target PSF choice, can be found in H24 and \citet{2024arXiv241005442C}, respectively. 

\section{Measurement of noise correlations}
\label{sec:noise_corr}

We investigate the noise fields from the lab data, providing an overview of the noise effects (Sec. ~\ref{subsec:noise-overview}), and then describing both the 2D (Sec. ~\ref{subsec:noise-2D}) and the azimuthally averaged (Sec. ~\ref{subsec:noise-1D}) power spectra.

\subsection{Overview}
\label{subsec:noise-overview}

There are two systematic observation errors produced by noise on a measurement: additive and multiplicative bias \citep{Heymans2006}. Given a true signal of $\gamma_{\rm true}$, noise and other biases produce a $\gamma_{\rm meas}$:
\begin{equation}\gamma_{\rm meas} = (1+m)\gamma_{\rm true}+c.\end{equation}

Additive bias $c$ represents an observed ``shear'' signal which occurs even in the absence of weak lensing on a target galaxy. This bias often occurs in the form of background noise. The second error is the multiplicative bias, $m$. This factor is a result of a real shear signal, and describes the difference between the measured signal and the true signal. 

The focus of this paper is on the additive shear bias due to noise, as propagated through the {\sc Imcom} framework. \newedit{Though both additive and multiplicative bias will need to be addressed in the \textit{Roman} WL analysis (and noise can contribute to both), the correlated noise is not statistically isotropic and therefore represents a particular concern for additive bias. Additionally, testing multiplicative bias levels requires propagating the noise through not just the shape measurement but also the shear calibration loop (which will likely be based on Metacalibration; \citealt{Sheldon2017}). We therefore defer analysis of multiplicative bias until after the \textit{Roman} shear calibration loop is ready for testing.}

The effect of noise depends on its spatial structure.
Any positive noise fluctuation --- even a ``pedestal'' (in the $u=0$ Fourier mode) --- will increase the signal from any object and thus lead to a bias in its flux (0$^{\rm th}$ moment of the image). Higher order moments of the noise fields introduce biases on different scales, that will change other types of measurements on our image.
The first moment $\langle x\rangle$ corresponds to an object's centroid position. 
This is related to the gradient of the noise field $\nabla I_{\rm noise}$, or in Fourier space, to $2\pi i u \tilde I_{\rm noise}$. The second moment of the image $\langle\,x^2\rangle$ is relevant to shape measurement. This is the signal we wish to pick out in order to precisely measure weak gravitational lensing. The second moment of the image is affected by the second derivative of the noise:
$\nabla_i\nabla_jI_{\rm noise}$, or $4\pi^2u^2\tilde I_{\rm noise}$ in Fourier space. 

While detailed predictions of the bias can be complicated (see, e.g., Y24), in all of these cases the biases can be related to the power spectrum of the noise (and its anisotropy), and higher-order moments of an object are sensitive to higher Fourier modes $u$. This motivates detailed examination of the noise power spectrum following image coaddition.

Y24 gives an analytic relation between the anisotropy of the noise power spectrum and the induced additive bias $c$ in the case of circular Gaussians; more realistic cases are treated numerically. But we still report the noise power spectrum, since it both provides the lowest-order summary statistic of the noise, and it is still true that it represents the contribution from noise at different scales to the noise bias contribution $c$. 

\subsection{2D noise power spectra}
\label{subsec:noise-2D}

The power spectrum $P(u,v)$ of a field $S$ is the 2D Fourier transform of its correlation function: 
\begin{equation}
P(u,v) = \int \xi(\Delta x, \Delta y) {\rm e}^{-2 \pi {\rm i}(u\Delta x + v\Delta y)} \,{\rm d}\Delta x\,{\rm d}\Delta y,
\end{equation}
where the correlation function is $\xi(\Delta x, \Delta y) = \langle S(x,y) S(x',y') \rangle$. Our data is discrete, so for this calculation we use a summation over the pixels in two dimensions in place of an integral. The power spectrum is defined over the first Brillouin zone of a square grid at output sampling scale $s_{\rm out}$, the square region where $|u|$ and $|v|$ are less than $1/(2s_{\rm out})$.

The T23-sims power spectra contained significant features caused by the repeated boundary effects from the edges of each block. This feature, a repetition at an exact period along \textit{x} and \textit{y} in real space, caused a straight line in Fourier space along the \textit{u} and the \textit{v} directions; it translates to a plus-sign shaped feature in the spectra. This is not true power at these spatial frequencies, but rather the result of window function leakage. To avoid this, we implemented an apodization scheme for the OU24-sims version of this work. Apodization is a common method for removing unphysical effects due to sampling from power spectra (see, e.g., \cite{Hamming1987}).
We apodize with a Tukey window, defined by 
\begin{equation}
\label{eq:tukeywin}\begin{cases} 
W_{\alpha}(n) = \frac{1}{2}\left[1 - \cos\left(\frac{2\pi n}{\alpha N}\right)\right], 
& 0 \leq n < \frac{\alpha N}{2} \\
W_{\alpha}(n) = 1, 
& \frac{\alpha N}{2} \leq n < \frac{N}{2} \\
W_{\alpha}(N-n) = W_\alpha(n), 
& 0 \leq n \leq \frac{N}{2},
\end{cases}
\end{equation}
where in this case $n$ is a pixel index, $N$ is the block length in pixels, and $\alpha$ parameterizes what fraction of the data is inside the cosine function, i.e. the rectangular part of the window has width \newedit{$(1-{\alpha})N$} \citep{Harris1978}. 
After testing different values of $\alpha$, we found that we could remove the periodic boundary effect using a narrow rectangular width, $\alpha=0.99$. 


The power spectrum is calculated using the Numpy Fast Fourier Transform (FFT) function \citep{Harris2020}. The lab noise frames are first converted from their raw intensity units of DN pix$^{-1}$ s$^{-1}$ to astrophysical units $\mu$Jy arcsec$^{-2}$ via
\begin{equation}\label{eq:LN_norm}
    \mathbb{S}_{\rm lab}[{\rm in~}\mu{\rm Jy\,arcsec}{^{-2}}] = \frac{hg}{s_{\rm in}^2 A_{\rm eff} } \mathbb{S_{\rm lab}}[{\rm in~DN~pix}{^{-1}}\,{\rm s}{^{-1}}]
\end{equation}
using each SCA's average gain $g$ as measured from the triplet test data using the photon transfer function technique (as implemented in \citealt{2020PASP..132g4504F}) and Planck's constant $h=662.6\,\mu\rm{Jy} \,\rm{cm}^2\,s$. The filter effective areas $A_{\rm eff}$ are calculated by integrating the effective area measurements\footnote{Effective areas are calculated based on: \url{https://github.com/GalSim-developers/GalSim/blob/releases/2.5/share/roman/Roman_effarea_20210614.txt} for consistency with OU24-sims. The triplet test calculations instead use effective areas from the early design phase of \textit{Roman} for consistency with the T23 simulations: $6051 {\rm cm}^2$ in J129 and $5978 {\rm cm}^2$ in H158.} over each filter's full bandpass, resulting in per-filter effective areas of:
\begin{equation}\label{eq:area}
    \int_{\nu} \frac{A_{\text{eff}}(\lambda) } \lambda \:\: d\lambda = 
    \begin{cases}
        7006\,  \text{ cm}^2, & \text{Y}106\\
        7111\,  \text{ cm}^2, & \text{J}129\\
        7340\,  \text{ cm}^2, & \text{H}158\\
        4840\,  \text{ cm}^2, & \text{F}184\\
        4654\,  \text{ cm}^2, & \text{K}213.
    \end{cases}
\end{equation}
For the K band only, we add an additional layer of white noise on top of the lab noise field to simulate the additional thermal background expected in this filter due to the warm telescope. Though in principle all filters have some amount of sky background, the count rate from zodiacal light and thermal background combined for Y, J, H, and F is $\leq 0.32$ per pixel (at minimum zodiacal light). In contrast, the much redder K band has an estimated background count rate of $4.65$ e pix$^{-1}$ s$^{-1}$.\footnote{Thermal background estimates \newedit{were taken from \url{https://roman.gsfc.nasa.gov/science/WFI_technical.html} in early 2024; values have since been updated but are not significantly different.}} 

We simulate the thermal background in the K band by using white noise frames, which were included in the OU24+\Pyimcom\ run, and are dimensionless with $\sigma_{\text{I}}=1 s^{-1}_{\text{in}}$. We rescale the white noise frames by 
\begin{equation}
    A_{WN} = \sqrt{\frac{\Delta\,B }{ t_{\rm exp}}} \frac{t_{\rm fr}}{g}
\end{equation}
to account for the additional thermal background $\Delta\,B=B_{\text{K213}}-B_{\text{PyImcom}}=4.64-0.38=4.27 $ e pix$^{-1}$ s$^{-1}$ over a total exposure of $t_{\rm exp}=139.8$s (Y24). This step occurs before conversion to physical units via Eq.~(\ref{eq:LN_norm}), so we also include the factor of $t_{\rm fr}/g$ to convert the noise frame to DN/frame, matching the lab noise frame units at this step in the process. 

Each lab noise block is next multiplied by the window function, Eq.~(\ref{eq:tukeywin}), and then normalized by the window effective area $A_{\rm win}$ and the input to output pixel scale ratio, $s_{\rm in}/s_{\rm out}$ to produce a power spectrum with a variance of $s_{\rm in}^{-2}$. For the unapodized (T23-sims) blocks, $A_{\rm win}=N^2$; apodized blocks (OU24-sims) instead have $A_{\rm win}=|W_{\alpha}(n)|^2$.

The 2D power spectrum of the lab noise is thus described by 
\begin{equation}P_{2D}(u,v) = \frac{s_{\rm out}^2}{A_{\rm win}^2}\left|
\sum_{j_x,j_y} \mathbb{S}_{\rm lab\,j_x,j_y} {\rm e}^{-2\pi{\rm i}s_{\rm out}(uj_x+vj_y)}
\right|^2,\end{equation}
 where $u$ and $v$ are integer multiples of $1/Ns_{\rm out}$ and $A_{\rm win}$ is the area of the region used to compute the power spectrum. 

The resulting 2D power spectra are binned into $8 \times 8$ bins, covering a range of $-12.8$ to $+12.8$ cycles arcsec$^{-1}$ in $u$ and $v$.

\begin{figure*}[ht]
    \centering
    \includegraphics[width=.8\linewidth]{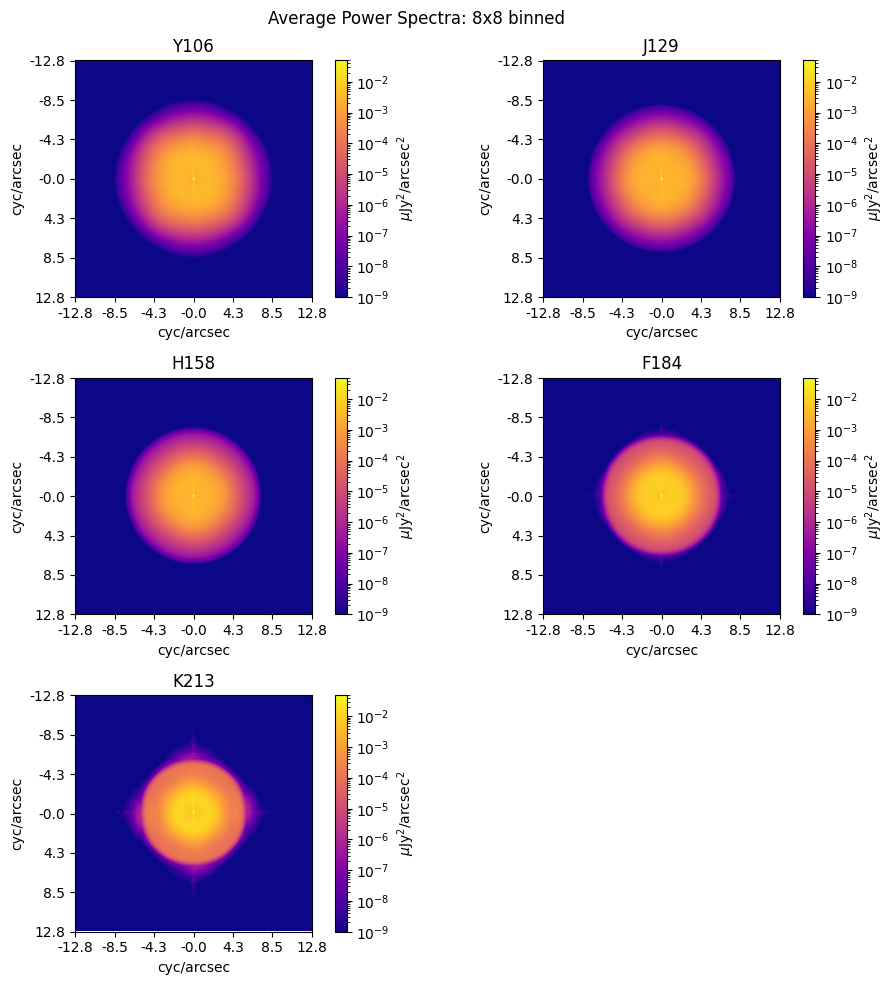}
    \caption{\label{fig:binned-spectra} The 2D averaged power spectrum of the lab noise fields in Y, J, H, F, and K bands, plotted on a logarithmic scale. The horizontal and vertical axes show the wave vector components ($u$ and $v$ respectively) ranging from $-12.8$ to $+12.8$ cyc/arcsec. The color scale shows the power $P(u,v)$ in units of $\mu$Jy$^2$ arcsec$^{-2}$. }
\end{figure*}

\begin{figure*}[ht]
    \centering
    \includegraphics[width=.8\linewidth]{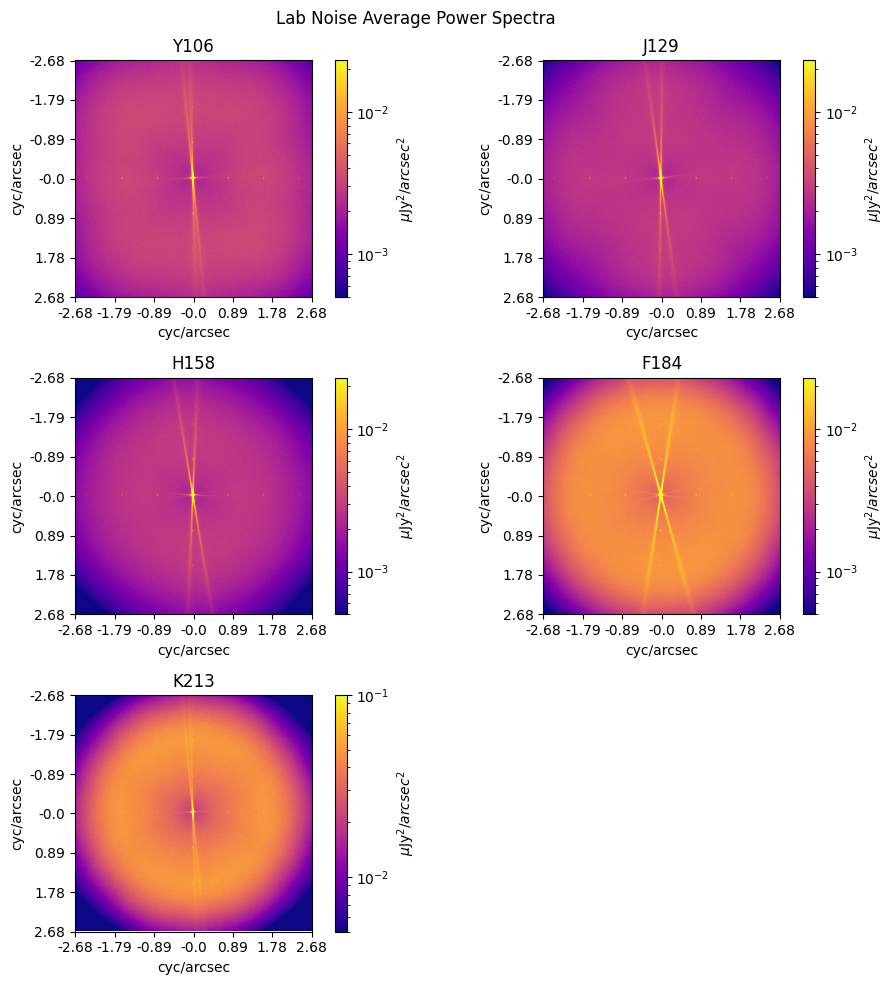}
    \caption{\label{fig:unbinned-spectra} Unbinned (linear scale) images of the central regions of the 2D lab noise power spectra. With the smaller scale for the power, the central features are more easily seen. The horizontal and vertical axes show the wave vector components ($u$ and $v$ respectively) ranging from $-2.68$ to $+2.68$ cyc/arcsec. The color scale shows power $P(u,v)$ in units of $\mu$Jy$^2$ arcsec$^{-2}$.}
\end{figure*}

Following the binning process, the power spectra are averaged over all 1296 (2304 for T23-sims) blocks to produce one 2D power spectrum for each filter. This complete process was done in the Y106, J129, H158, F184, and K213 bands, which are all under consideration for the HLIS survey design. The resulting binned power spectra can be seen in Fig.~\ref{fig:binned-spectra}. Additionally, we produce an unbinned averaged power spectrum in each band in  Fig. \ref{fig:unbinned-spectra}; we plot only the central \newedit{$\sim 5$ cyc/arcsec} regions of these spectra, using the higher resolution to display particular features. Minimum and maximum values of the power spectra are reported in Table \ref{tab:spectra_minmax}. 

Real noise has components of both white noise and $1/f$ noise, as well as various other components, which contribute less to the overall noise variance. 
Therefore, we expect to see similar features from the simulated white and $1/f$ power spectra from Y24 in the power spectra of the lab data. Any outstanding, unfamiliar features require further analysis. 

Because our output PSF is larger than the input PSF in real space, we expect high wavenumbers to have reduced power. This is a feature we clearly see in all bands; power is maximized in the central regions of the plot, and falls off smoothly. 

Additionally, in Figure \ref{fig:unbinned-spectra} we see that in all bands, power is not peaked in the center and monotonically decreasing as we might naively expect for a Gaussian PSF. This is due to the process of constructing an oversampled image from an undersampled one. As multiple undersampled images are combined together, each Fourier mode being reconstructed aliases to other nearby Fourier modes. This effect is treated in detail in \citet{Lauer1999a}. For the \textit{Roman} coadded noise power spectra, the aliasing of modes into other modes causes noise to be amplified at all frequencies except $u=0$, since that mode does not alias to any others. We observe a local minimum in the region around $(u,v)=(0,0)$ and an increase in noise afterwards (before the eventual Gaussian decrease), with the overall effect of a ``donut'' shape.

K213 has the largest noise, which makes sense since it is the reddest band and especially when including the estimated thermal background (as we do here). K213 and F184 also display a sharper cutoff between the high and low power regions in their spectra. This cutoff is in general specified by the PSF, which in the other bands is dominated by the Gaussian but in these bands specifically is diffraction-limited because of the redder bandpasses. 
To test whether the cutoff comes from the diffraction limit, we calculate the expected radius of the central region for the Airy disk,
\begin{equation}
    R_{\lambda} = \frac{D}{\lambda}\frac{s_{\rm out}}{206265~\rm arcsec} N,
\end{equation}
using the central wavelength $\lambda=2.13\,\mu\rm{m}$ for K213 and $\lambda=1.84\,\mu\rm{m}$ for F184, the telescope diameter $D=2.37$m, our $s_{\rm out}=0.039''$ pixel scale, and an FFT array dimension of $N=2496$ pixels.
We find $R_{\rm K213}=525$ pixels and $R_{\rm F184} = 607$ pixels, which are approximately the radii of the bright central regions in those bands.
Because the \Pyimcom\ linear algebra knows there should not be any sky signal outside the band limit, modes outside this limit are nulled out, causing the hard cutoff in these bands. In contrast, in the bluer bands the Fourier-space target PSFs flow smoothly to zero before their diffraction limit is reached, so the steep cutoff is avoided.

We also observe some deviations from perfect circular Gaussian shapes in the spectra, depending on the band. This is most likely due to different levels of average observation coverage; Y106 has the lowest overall coverage (H24), and its spectrum shows the most distinct non-circular central region. 

In order to maximize coverage in each filter, the Reference Survey proposes two passes with unique roll angles in each filter \cite{Spergel2015}. It is known that the \textit{Roman} detectors and electronics produce a striping feature in the noise fields; this can be seen clearly in Fig.~\ref{fig:refcor}. The X feature seen in the middle of the power spectra comes from the striping pattern in the noise fields at the two distinct roll angles. In Y106 and K213, the X is very narrow and hard to resolve because the roll angles are not very far apart ($\approx 5$ degrees).
The minimum roll angle separation is a future question for the survey definition committee.

The center of the X feature also repeats along the \textit{u} and \textit{v} axes along some grid. This can be seen more clearly in Figure \ref{fig:unbinned-spectra}. Inspecting the spectra reveals $\Delta u = 78$\,pix = $0.8$ cyc/arcsec = $1.25$ arcsec, and the same in the \textit{v} direction. The feature repeats every 1.25 arcsec, which is the size of the postage stamps.  We conclude that the grid of spots is caused by the postage stamps. 

\begin{table}
    \centering
    \begin{tabular}{c|c|c}
    \hline
        Filter  & $P_{\text{min}}~[\mu{\rm Jy}^2/{\rm arcsec}^2]$ & $P_{\text{max}}~[\mu{\rm Jy}^2{\rm/arcsec}^2]$ \\
        \hline\hline
        Y106 &  $6.09 \times 10^{-17}$  &  $5.51$  \\
J129 &  $4.64 \times 10^{-17}$  &  $4.96$  \\
H158 &  $3.80 \times 10^{-17}$  &  $5.11$  \\
F184 &  $1.12 \times 10^{-16}$  &  $12.55$  \\
K213 &  $5.10 \times 10^{-16}$  &  $12.57$  \\

\hline
    \end{tabular}
    \caption{Minimum and maximum 2D power of noise slope in each filter.}
    \label{tab:spectra_minmax}
\end{table}

\subsection{Azimuthally averaged power spectra}
\label{subsec:noise-1D}

\begin{figure}[h]
    \centering
    \includegraphics[width=0.9\linewidth]{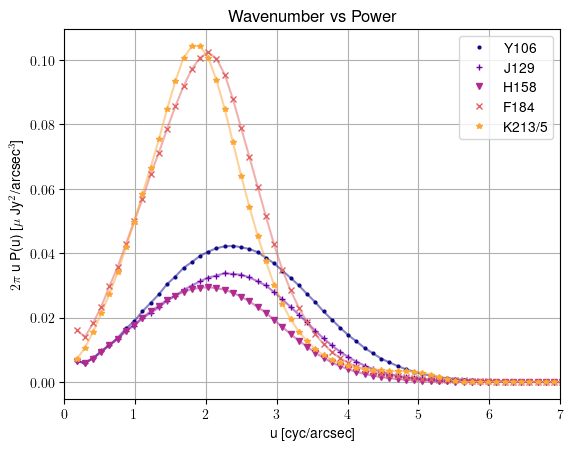}
    \caption{\label{fig:OU24-1dspectra} The 1D azimuthally-averaged power spectra of lab noise fields measured in each band and coadded with \Pyimcom\,. Note that K213 is divided by 5 on this plot, as its maximum $P_{\rm max}=1.28\times10^{-5}$ is an order of magnitude higher than all the other bands, making the figure hard to read if K is plotted to scale.}
\end{figure}

The 1D power spectra are built from the 2D versions described in the previous section. The 2D image is sectioned into concentric circular bins increasing in size outward from the center. We compute the average power in each azimuthal ring, and plot relative to the relevant scale (arcseconds). The resulting power spectra are shown in Figure~\ref{fig:OU24-1dspectra}.

In generating these averaged spectra, it is relevant to note that the survey strategy (combining multiple dithered exposures in each filter) generates uneven coverage of each region of the sky. Gaps between individual chips in the focal plane's design also contribute to this uneven coverage map. Figure 1 of H24 shows the coverage map for the full 48$\times$48 arcminute region considered in the T23-sims run of \Imcom\,; though this is not an accurate coverage map for the OU24-sims data used in this work, the survey strategy is not significantly different so the overall amount of coverage in each filter is similar. Observation coverage maps for OU24 will be included in the data release note, which is currently in prep. from the LSST-DESC + \textit{Roman} team. Regions with lower overall sky coverage will have higher levels of noise (see Appendix \ref{app:T23}).


Real detector noise contains elements of both correlated and uncorrelated noise, so we expect to see behaviors common to both in the power spectra. The overall shape of the 1D power spectra seen in Figure \ref{fig:OU24-1dspectra} resembles a superposition of white noise and $1/f$ noise (see Figure 4 of Y24), as expected.

The power spectra are highly suppressed on both very small and very large scales. This is related to the output PSF size and shape, just as in the 2D spectra. Noise correlations on the smallest scales should be close to zero, bringing the left-hand tails of the spectra down towards zero. The right-hand tails of the spectra are similarly suppressed because the output PSF is narrower in Fourier space (larger in real space) than the input target PSF. The tails of the spectra approach zero at the edge of the bright central regions of the 2D spectra, around 6 cyc/arcsec. For the K band, the transition down to zero is slightly less smooth; because of the larger amount of noise, the spectrum has higher power farther out in $u$ space and then decreases sharply to zero when the diffraction limit is reached, around $u=D/1.03\lambda=5.2$.

White noise is smoothed by the Gaussian output PSF, so we expect the overall Gaussian envelope observed in these power spectra. In the simplistic case of white noise smoothed by a Gaussian of full width at half maximum $\theta_{\rm FWHM,sm} = (\theta_{\rm FWHM,out}^2 - \theta_{\rm FWHM,in}^2)^{1/2}$, the power spectrum should be $P(u) \propto e^{-\theta_{\rm FWHM,sm}^2u^2/(2\ln 2)}$. Then even though $P(u)$ peaks at $u=0$, the power per unit 1D wave number $2\pi u P(u)$ peaks at $u_{\rm peak}=\sqrt{\ln 2}\,/\pi\theta_{\rm FWHM,sm}$. Using the diffraction limit 1.03$\lambda/D$ for $\theta_{\rm FWHM,in}$ and the smoothed Gaussian $\theta_{\rm FWHM,out}$ we calculate $\theta_{\rm FWHM,sm}$ for each filter and find that the true $u_{\rm peak}$ is systematically higher (about twice as large) than we expect from this calculation.

Figure $\ref{fig:OU24-1dspectra}$ shows clearly that the H and J filters are ideal for minimal noise correlations. We can also see that F and K filters are probably not ideal for shape measurement due to very large amounts of noise correlations. This will be shown further in Section \ref{sec:noise_bias}.



\section{Estimation of additive noise bias}\label{sec:noise_bias}

In addition to measuring the noise properties, we wish to ensure that these signal components do not potentially contaminate or bias the cosmological parameters estimated from weak lensing observations. In this section, we will estimate the noise bias on our injected stars and galaxies, and calculate the shape-shape correlation function $\xi_\pm(\theta)$ induced by noise bias. 

Since the \newedit{angular} power spectrum depends on the square of measured signal, the measured \newedit{angular} power spectrum is
\begin{equation}
C_\ell^{\gamma_{\rm obs} \gamma_{\rm obs}} = (1+m)^2 C_\ell^{\gamma_{\rm true}\gamma_{\rm true}} + C_\ell^{cc}
\end{equation}
where $C_\ell^{cc}$ is the power spectrum contribution from additive bias $c$. 
The \textit{Roman} SRD provides a requirement on the power spectrum of $c$, which has been converted to correlation function space in Y24.

To start, we convert the fields of injected objects and the lab noise fields into physical units. The injected stars and galaxies are normalized to unit flux, so before adding in the noise fields we multiply them by conversion factor
\begin{equation}\label{eq:normsrc}
    A_{\text{source}} = \frac{1}{s^2_{\text{in}}} F_{0,AB} 10^{-0.4\,m_{AB}}\,\, ,
\end{equation}
where $F_{0,AB} = 3.631\times10^9$ is the usual astronomical magnitude scaling to units of microjansky and we set $m_{AB}=23.0$, 23.5, 24.0, or 24.5, spanning the range of magnitude for \textit{Roman} shape measurement requirements. Similarly, we convert the lab noise fields from DN/frame to $\mu$Jy/arcsec$^2$ via
\begin{equation}\label{eq:normnoise}
    A_{\text{\rm noise}} =  \frac{g}{t_{\text{fr}}} I_{\text{tot}} = \frac{g}{t_{\text{fr}}\, s^2_{\text{in}}} \int_{\nu} \frac{h\nu}{A_{\text{eff}}(\nu)} d\nu,
\end{equation}
where $t_{\text{fr}} = 3.08 $s is the frame time, $g=1.458$ e/DN is the gain, and $I_{\text{tot}}$ is the total intensity in the image, given by the integral over frequencies of the flux times effective area of the filter $A_{\text{eff}}(\nu)$ (defined in \ref{subsec:noise-2D}). As with the power spectrum analysis, we include a layer of white noise in the K band to account for additional thermal background. White noise is rescaled to simulate the estimated thermal background, converted to physical units to match the lab noise, and added to the lab noise frame  (see \S \ref{subsec:noise-2D} for details).

\begin{figure*}[ht]
    \centering
    \includegraphics[width=.8\linewidth]{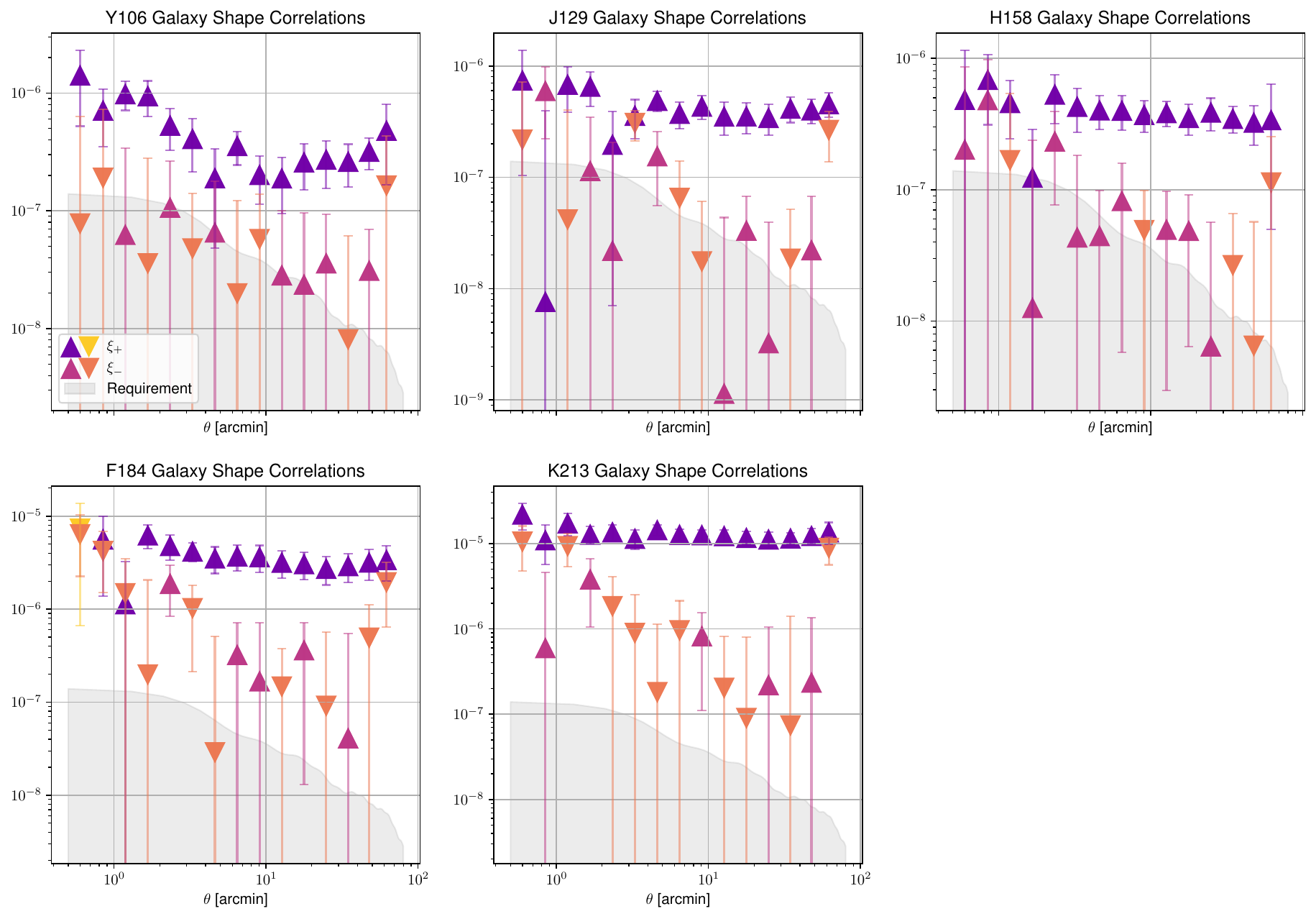}
    \caption{\label{fig:ggcorrelations_g} Magnitude of the correlation function of noise bias on shapes ($\xi_{+/-}(\Delta\,g_1, \Delta\,g_2)$) for {\sc GalSim} injected galaxies with magnitude ${\rm m}_{\rm AB}=24$. The grey shaded region indicates the approximate required value for $\xi_+$, calculated based on the additive bias error budget in the SRD. Upward-pointing triangles represent positive-valued $\xi$s while downward-pointing triangles represent negative-valued $\xi$s. Purple and yellow triangles represent the values of $\xi_+$ while pink and orange are $\xi_-$. }
    
    \centering
    \includegraphics[width=.8\linewidth]{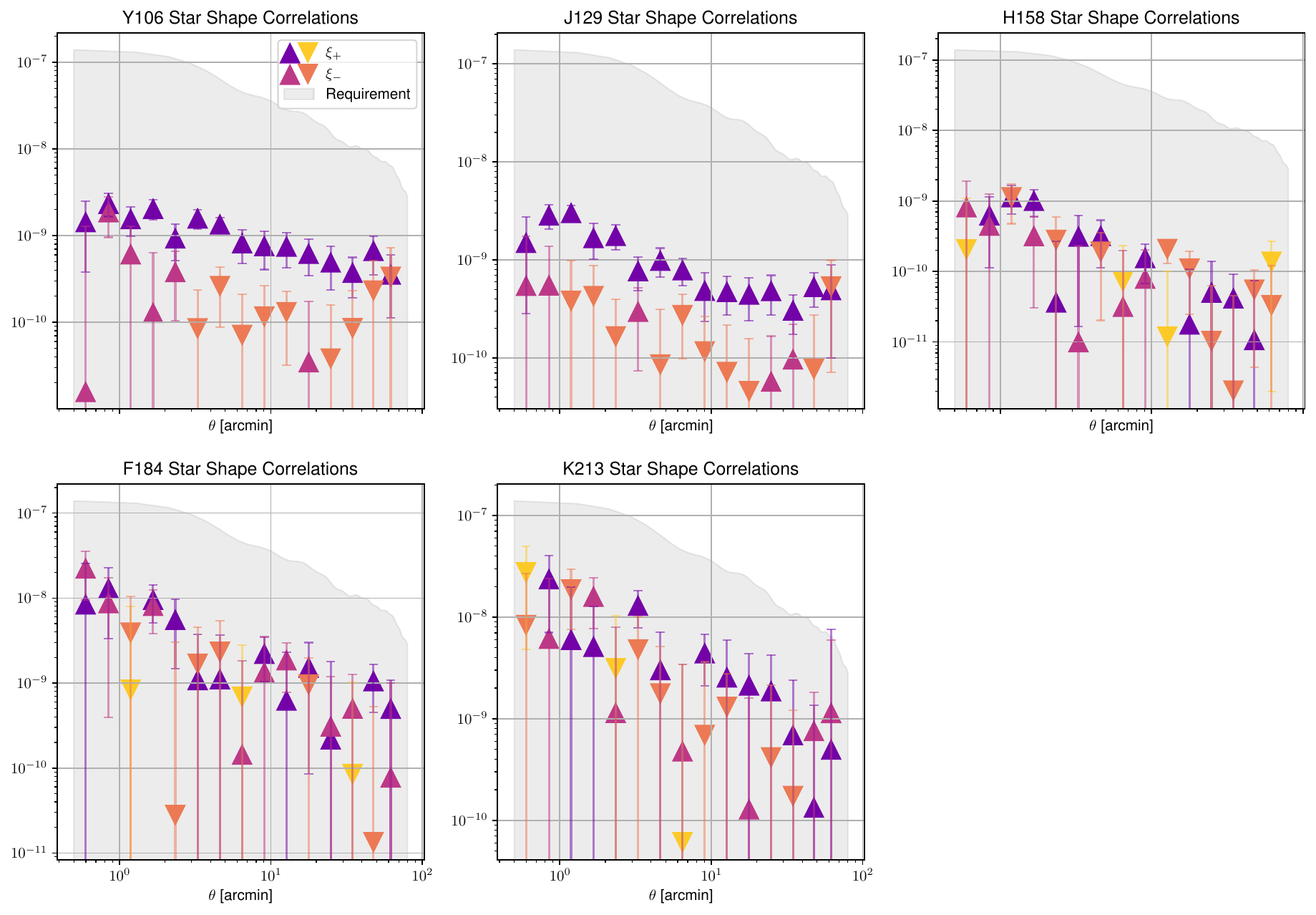}
    \caption{\label{fig:ggcorrelations_s} This figure shows the magnitude of the correlation function of noise bias on shapes ($\xi_{+/-}(\Delta\,g_1, \Delta\,g_2)$) for {\sc GalSim} injected stars with magnitude ${ m}_{\rm AB}=24$. The grey shaded region indicates the approximate required value for $\xi_+$, calculated based on the additive bias error budget in the SRD. Upward-pointing triangles represent positive-valued $\xi$s while downward-pointing triangles represent negative-valued $\xi$s. Purple and yellow triangles represent the values of $\xi_+$ while pink and orange are $\xi_-$.}
\end{figure*}

Next we use these normalized lab noise images $I_{\rm noise}$ to construct ``positive'' and ``negative'' noisy images of the sources,
\begin{equation}
    I_{\pm {\rm noise}} = I_{\rm src} \pm I_{\rm noise},
\end{equation}
and measure their shapes using the {\tt galsim.hsm} adaptive moments module in the {\sc GalSim} package \citep{2003MNRAS.343..459H, 2015A&C....10..121R}. Object shapes are parameterized into shear components $(g_1, g_2)$, which are defined by the reduced shear $|g|=(a-b)/(a+b)$ ($a$ and $b$ being the major and minor axes) and the position angle $\beta$ via
\begin{equation}
    g_1 = |g|\cos(2\beta) ~~~{\rm and}~~~
    g_2 = |g|\sin(2\beta).
\end{equation}

Finally, we combine each shear component measured from the three different source images (positive and negative noise realizations, and noiseless) via
\begin{equation}
    \Delta g_{1,2} = \frac{g_{1,2}(I_\text{src} + I_{\text{\rm noise}}) + g_{1,2}(I_\text{src} - I_{\text{\rm noise}})}{2} - g_{1,2}(I_\text{src}),
\end{equation}
giving the additive noise bias of each shear component, $\Delta\,g_{1,2}$. \newedit{The mean additive noise bias of galaxy shear $\big<\Delta\,g_{1,2}\big>$ is reported in Table \ref{tab:meandeltag}. We can see directly that there is a significant offset caused by noise in the shear measurements. As we run simulations with larger mosaics in the future, we will be able to study the coherence scale of this large-scale shear and determine whether it is in the range of angular scale $\ell$ for the \textit{Roman} requirements.}

\begin{table}
    \centering
    \begin{tabular}{c|c|c}
    \hline
        Filter  & $\Delta\,g_1$ & $\Delta\,g_2$ \\
        \hline\hline
Y106 & $4.71\times 10^{-4}$ & $-3.29\times 10^{-4}$ \\
J129 & $6.36\times 10^{-4}$ & $-1.91\times 10^{-4}$ \\
H158 & $6.13\times 10^{-4}$ & $-1.16\times 10^{-4}$ \\
F184 & $1.76\times 10^{-3}$ & $-2.54\times 10^{-4}$ \\
K213 & $3.32\times 10^{-3}$ & $-1.00\times 10^{-3}$ \\
\hline
    \end{tabular}
    \caption{\newedit{Mean shear bias from noise on galaxies of magnitude $m_{\rm AB}=24$ in each filter. The presence of a mean shear offset is likely the cause of the nearly-flat correlation functions in Figure~\ref{fig:ggcorrelations_g}. Notably, for all filters, the mean galaxy shear bias is larger than the SRD requirement on RMS additive shear variance per component $c_{\rm req}=7.0\times10^{-5}$ for $1.5\leq \log_{10}\ell\leq2.5$. This is a first indication that noise bias corrections (at the level of $\sim 1$ order of magnitude) will be necessary.}}
    \label{tab:meandeltag}
\end{table}

We use \Treecor\ \citep{Jarvis2004} to calculate the correlation functions $\xi_{+,-}$ between the two components of additive shear, shown in Figure~\ref{fig:ggcorrelations_g} (for a sample of injected analytic galaxies, with an exponential profile and with half-light radius log-distributed between 0.125 and 0.5 arcsec, and with intrinsic ellipticities drawn uniformly in the region of the $(g_1,g_2)$ with $|g|<0.5$) and Figure~\ref{fig:ggcorrelations_s} (for injected stars), in each observing band.

To compare the measured correlation function with the error budget for \textit{Roman}, we estimate the requirement on $\xi_+$ following Y24 and \cite{Givans2022}. The spin-0 correlation $\xi_+$ is approximated by
\begin{equation}
    \xi_+(\theta)\simeq\sum_{\ell\, \rm{bins}}2\gamma_{\rm rms}^2\overline{J_0(\ell\theta)},
\end{equation}
a Riemann sum over bins of multipoles $\ell$ where $\gamma_{\rm rms}$ is the allowed root-mean-square shear per component and $\overline{J_0}$ the Bessel function of the first kind, averaged over $\ell$. Using the required variance per component \newedit{of additive shear bias} from the \textit{Roman} SRD, we calculate and plot our approximate $\xi_+$ requirement alongside the correlation functions in Figures \ref{fig:ggcorrelations_g} and \ref{fig:ggcorrelations_s}.

Figure \ref{fig:ggcorrelations_s} shows that the contribution of additive noise bias to the shape-shape correlations for stars is well below the mission requirements. This was the case using the T23-sims+\Imcom\, data as well (shown in \ref{fig:T23-ggcorrelations}). However, we find for injected galaxies the situation is significantly worse. For the injected galaxies, the average correlation function value lies around $10^{-6}$ in bands with high coverage, and $10^{-5}$ in bands with less complete coverage. This is 1--2 orders of magnitude larger than the estimated mission requirement. Figure \ref{fig:ggcorrelations_g} shows this clearly.

It is expected that noise correlations matter more for extended objects. We have normalized both sets of sources to the same magnitude $m_{\rm AB}=24$. Galaxies are bigger objects, so the magnitude gets spread out over a larger area and results in a smaller surface brightness. This makes our objects fainter for detection and measurement purposes. 

To test the limits of this effect, we performed all of the analysis in this section with injected sources renormalized to magnitudes $23, 23.5, 24$ and $24.5$. This will indicate how faint of sources are reliably measured in each band. We additionally break the field of object detections into 3 scale ranges, defined by $\frac{1}{2}\theta_{\rm SCA}$ and $2\theta_{\rm SCA}$. The mean values of $\xi_{\pm}(\Delta\,g_1, \Delta\,g_2)$ for galaxies over this range of scales and magnitudes are included in Table \ref{tab:correlations_mags_g}. 

\begin{figure}[ht]
    \centering
    \includegraphics[width=3.2in]{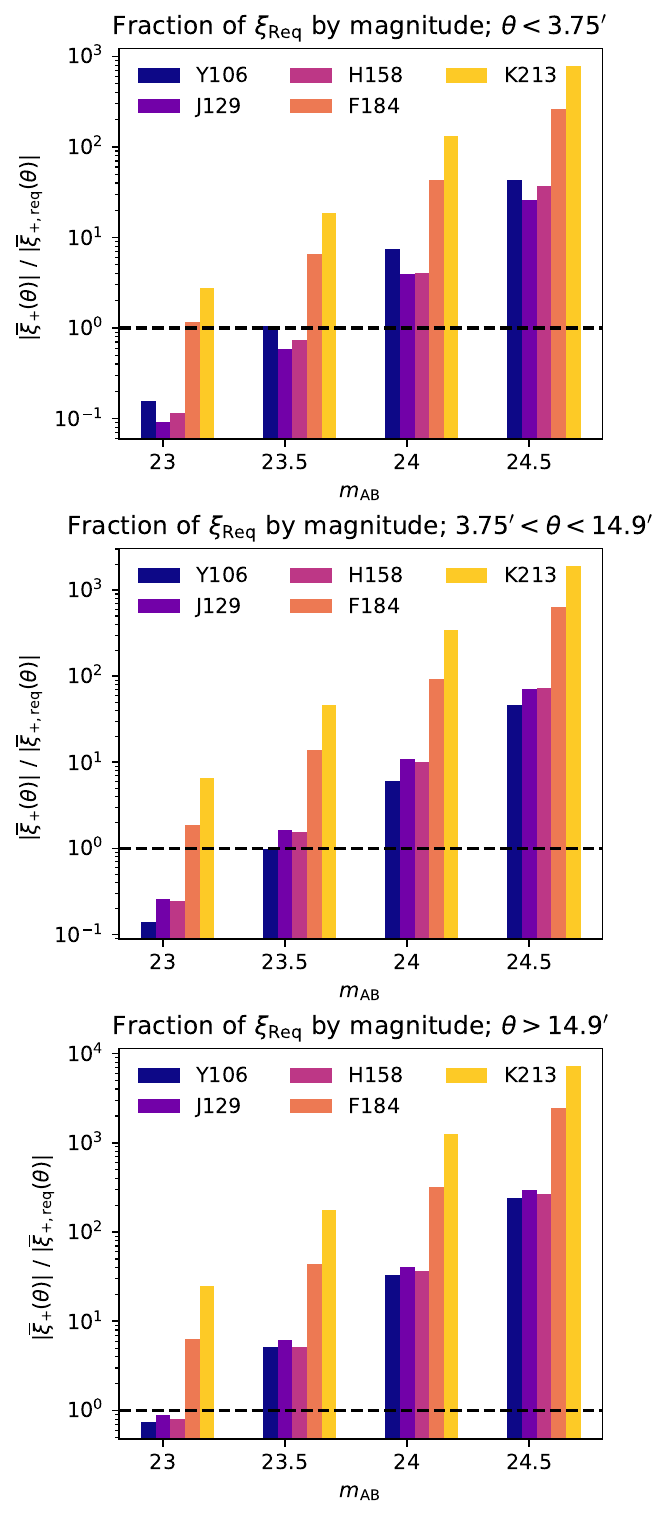}
    \caption{Mean $\xi_+$ as a fraction of the mean required $\xi_+$, over 3 scale ranges defined by ($\theta_{\rm SCA}-1/2\theta_{\rm SCA}$, $\theta_{\rm SCA}+2\theta_{\rm SCA}$). The horizontal dashed line shows the requirement cutoff; bars below the line are within it, and bars above the line will require mitigation.}
    \label{fig:corr_reqs}
\end{figure}

Additionally, Figure \ref{fig:corr_reqs} shows the ratio of the measured $\xi_{\pm}$ means to the mean of the required $\xi_+$ from the Roman SRD. We calculated the mean required $\xi_+$ in each range:
\begin{equation}
       \xi_{+, {\rm req}}= 
       \begin{cases}
        1.13\times10^{-7}, & \theta<3.75' \\
        3.84\times10^{-8}, & 3.75'<\theta<14.9' \\
        9.71\times10^{-9}, & \theta>14.9'. \\
        \end{cases}
\end{equation}
Comparing these values with the mean $\xi_+$ for different magnitudes over the same scale ranges, we see that noise biases are within requirements only for $m_{\rm AB}<23$ on all scales, and up to $m_{\rm AB}=24$ for a few cases (on small scales or for bands with high coverage). Additionally we see that the K band has the largest noise biases, which is expected due to the much higher thermal background.

\begin{table*}
\centering
\begin{tabular}{|l|c|c|c|c|}
\hline
Band  &  $m_{\rm AB}$ & $\langle\,|\xi_{+}(\theta<3.75')|\rangle$ & $\langle\,|\xi_{+}(3.75'<\theta<14.99')|\rangle$ & $\langle\,|\xi_{+}(14.99'<\theta)|\rangle$ \\
\hline\hline
\multirow{4}{*}{Y} & 23 & \((1.76 \pm 0.92) \times 10^{-8}\) & \((5.43 \pm 2.78) \times 10^{-9}\) & \((7.33 \pm 3.33) \times 10^{-9}\) \\
\cline{2-5}
 & 23.5 & \((1.18 \pm 0.63) \times 10^{-7}\) & \((3.74 \pm 1.78) \times 10^{-8}\) & \((4.97 \pm 2.31) \times 10^{-8}\) \\
\cline{2-5}
 & 24 & \((8.37 \pm 3.77) \times 10^{-7}\) & \((2.36 \pm 1.10) \times 10^{-7}\) & \((3.20 \pm 1.49) \times 10^{-7}\) \\
\cline{2-5}
 & 24.5 & \((4.87 \pm 2.21) \times 10^{-6}\) & \((1.79 \pm 0.68) \times 10^{-6}\) & \((2.33 \pm 0.96) \times 10^{-6}\) \\
\hline\hline
\multirow{4}{*}{J} & 23 & \((1.05 \pm 0.83) \times 10^{-8}\) & \((9.78 \pm 2.91) \times 10^{-9}\) & \((8.77 \pm 2.43) \times 10^{-9}\) \\
\cline{2-5}
 & 23.5 & \((6.58 \pm 5.56) \times 10^{-8}\) & \((6.35 \pm 1.79) \times 10^{-8}\) & \((5.99 \pm 1.50) \times 10^{-8}\) \\
\cline{2-5}
 & 24 & \((4.43 \pm 3.15) \times 10^{-7}\) & \((4.14 \pm 1.05) \times 10^{-7}\) & \((3.96 \pm 1.07) \times 10^{-7}\) \\
\cline{2-5}
 & 24.5 & \((2.97 \pm 1.98) \times 10^{-6}\) & \((2.71 \pm 0.72) \times 10^{-6}\) & \((2.88 \pm 0.70) \times 10^{-6}\) \\
\hline\hline
\multirow{4}{*}{H} & 23 & \((1.32 \pm 0.93) \times 10^{-8}\) & \((9.28 \pm 2.94) \times 10^{-9}\) & \((7.81 \pm 3.49) \times 10^{-9}\) \\
\cline{2-5}
 & 23.5 & \((8.38 \pm 5.34) \times 10^{-8}\) & \((6.00 \pm 1.90) \times 10^{-8}\) & \((5.04 \pm 2.08) \times 10^{-8}\) \\
\cline{2-5}
 & 24 & \((4.54 \pm 3.00) \times 10^{-7}\) & \((3.91 \pm 1.05) \times 10^{-7}\) & \((3.52 \pm 1.36) \times 10^{-7}\) \\
\cline{2-5}
 & 24.5 & \((4.22 \pm 1.99) \times 10^{-6}\) & \((2.83 \pm 0.63) \times 10^{-6}\) & \((2.57 \pm 0.80) \times 10^{-6}\) \\
\hline\hline
\multirow{4}{*}{F} & 23 & \((1.33 \pm 0.63) \times 10^{-7}\) & \((7.23 \pm 2.67) \times 10^{-8}\) & \((6.11 \pm 2.44) \times 10^{-8}\) \\
\cline{2-5}
 & 23.5 & \((7.51 \pm 4.18) \times 10^{-7}\) & \((5.35 \pm 1.78) \times 10^{-7}\) & \((4.25 \pm 1.68) \times 10^{-7}\) \\
\cline{2-5}
 & 24 & \((4.90 \pm 2.88) \times 10^{-6}\) & \((3.54 \pm 1.13) \times 10^{-6}\) & \((3.08 \pm 1.10) \times 10^{-6}\) \\
\cline{2-5}
 & 24.5 & \((2.98 \pm 1.57) \times 10^{-5}\) & \((2.48 \pm 0.70) \times 10^{-5}\) & \((2.38 \pm 0.71) \times 10^{-5}\) \\
\hline\hline
\multirow{4}{*}{K} & 23 & \((3.15 \pm 1.07) \times 10^{-7}\) & \((2.51 \pm 0.51) \times 10^{-7}\) & \((2.38 \pm 0.45) \times 10^{-7}\) \\
\cline{2-5}
 & 23.5 & \((2.10 \pm 0.63) \times 10^{-6}\) & \((1.77 \pm 0.33) \times 10^{-6}\) & \((1.72 \pm 0.29) \times 10^{-6}\) \\
\cline{2-5}
 & 24 & \((1.48 \pm 0.45) \times 10^{-5}\) & \((1.31 \pm 0.20) \times 10^{-5}\) & \((1.22 \pm 0.25) \times 10^{-5}\) \\
\cline{2-5}
 & 24.5 & \((8.83 \pm 1.98) \times 10^{-5}\) & \((7.31 \pm 1.22) \times 10^{-5}\) & \((7.02 \pm 1.11) \times 10^{-5}\) \\
\hline
\end{tabular}
\caption{Average values of noise bias shape correlations for injected galaxies of different magnitudes.}
\label{tab:correlations_mags_g}
\end{table*}

\section{Summary and Discussion}
\label{sec:discussion}

One of the largest challenges to the execution of a weak gravitational lensing survey with the \textit{Nancy Grace Roman Space Telescope} is control of systematic errors. 
In previous works (H24, Y24), simulated noise fields have been used to estimate the impacts of noise on measurements of galaxy shapes. In this work, we utilize real detector noise data taken in the lab to obtain the most realistic estimates of noise-induced shape measurement bias yet.

Laboratory noise frames were pre-processed following the steps in Section \ref{subsec:preprocess} to simulate the steps that will be taken in processing of \textit{Roman} images. The noise frames studied in this work are a combination of ``slope'' images of the detector read noise and estimated sky background in each filter.

As object shape is directly related to the second derivative of the image in real space (see \ref{sec:noise_corr}), we use power spectra of the noise frames to investigate shape bias. We analyze the full 2D power spectrum and an azimuthally averaged 1D spectrum. Additionally, we estimate the amount of additive noise bias by combining the lab noise frames with simulated grids of stars and galaxies (see \ref{sec:noise_bias}) and measuring the shape-shape correlation function. Significant results of these analyses include:

\begin{itemize}
\item \textit{Roman} detector noise power spectra show behaviors common to white and \textit{1/f} correlated noise.
\item The spectra also show a large X shape caused by correlated noise stripes at multiple roll angles of images. Removal of correlated noise is a future goal for the \textit{Roman} image processing pipeline development team. Several groups are working towards this end, including \citet{Rauscher2022} by using reference pixel corrections, and an ongoing work by this group developing a method to de-stripe images at the image combination stage using the two roll angles. 
\item Features caused by choices made during coaddition are also observed, including from block and postage stamp size and boundary conditions. In this paper we implemented a window function which efficiently removed some of these effects from the power spectra; we recommend the use of a similar method in future analyses.
\item The magnitude of noise biases on the shape-shape correlations is below the mission requirements for stars but \textit{not for galaxies}, with a slight dependence on magnitude. Galaxies of magnitudes $m_{\rm AB}\gtrsim23.5$ have correlations 1--2 orders of magnitude larger than mission requirements. Since the \textit{Roman} weak lensing analysis is planned to include galaxies up to $m_{\rm AB}=24.5$, we emphasize the necessity of developing noise de-biasing techniques for weak lensing.
\item Finally, we do not recommend the use of the K213 band for shape measurement, due to the larger thermal background and hence larger noise biases. The use of K213 should still be considered for other purposes in a weak lensing analysis, such as for photometric redshift calibration in the deep fields, but this is outside the scope of this paper. 

\end{itemize}

The conclusions above paint a clear picture of detector systematics that can be expected to appear in the \textit{Roman} HLIS data set. By analyzing true detector noise through the \Pyimcom\, pipeline, we can see where we are in the pursuit of achieving the mission requirements level, and lay out a path forwards towards unbiasing the \textit{Roman} images. Future work will include the development of an image destriping algorithm and inclusion of self-Poisson noise in the simulated sky background.

\section*{Acknowledgements}

We thank Ed Wollack, Jahmour Givans, and Chaz Shapiro for useful feedback on this project.

This paper is based on data acquired at the Detector Characterization Laboratory at NASA Goddard Space Flight
Center.

Computations were performed on the Pitzer cluster at the Ohio Supercomputer Center \cite{Pitzer2018}.

K.L., E.M., C.H., and K.C. received support from the National Aeronautics and Space Administration, under subaward AWP-10019534 from the Jet Propulsion Laboratory. C.H. additionally received support from the David \& Lucile Packard Foundation award 2021-72096. M.Y. and M.T. were supported by NASA under JPL Contract Task 70-711320, ``Maximizing Science Exploitation of Simulated Cosmological Survey Data Across Surveys.'' M.T. was supported by the ``Maximizing Cosmological Science with the Roman High Latitude Imaging Survey'' Roman Project Infrastructure Team (NASA grant 22-ROMAN11-0011).

\section*{Data Availability}

\bibliography{mainbib}{}
\bibliographystyle{aasjournal}

\appendix
\section{Noise Correlations}\label{app:T23}

In this Appendix, we present the results of the power spectrum and noise shape correlations analysis from a previous version of noise realizations and image simulations. 

The image simulations used here are those of \citet{Troxel2023} (T23-sims), which are analyzed in depth in \citet{Hirata2023} (H24) and \citet{Yamamoto2023} (Y24). We refer the interested reader to those works for details on the simulations and their results. The T23 simulations here are combined with noise realizations from the Triplet tests of the \textit{Roman} detectors, further described in \S \ref{sec:labdata}. 

The work in this appendix was meant to go along with H24 and Y24, but a new set of noise realizations was generated in the middle of that analysis and so the work was postponed to await the OpenUniverse 2024 simulations used in the majority of this work. However, for the sake of completeness, we include our analysis of the T23-sims lab noise images in this Appendix. We emphasize that the results shown in this Appendix are not as realistic as the results achieved in the rest of this paper. Additionally, the image coaddition had already been completed in the Y and F bands before this work was commenced, so for the T23-sims analysis of lab noise we only have results from H and J bands.

\subsection{2D Spectra}

Following the procedure as outlined in \S \ref{sec:noise_corr}, we calculated the noise power spectra ($8\times8$ binned) in the J and H bands. Though the procedure is the same, we note that the normalization of the lab noise data is slightly different. The values of $A_{eff}(\nu)$ used for the Triplet test calculations were based on a previous measurement of effective areas. These values can be found in Y24.

\begin{figure*}[hb]
    \centering
    \includegraphics[width=6.5in]{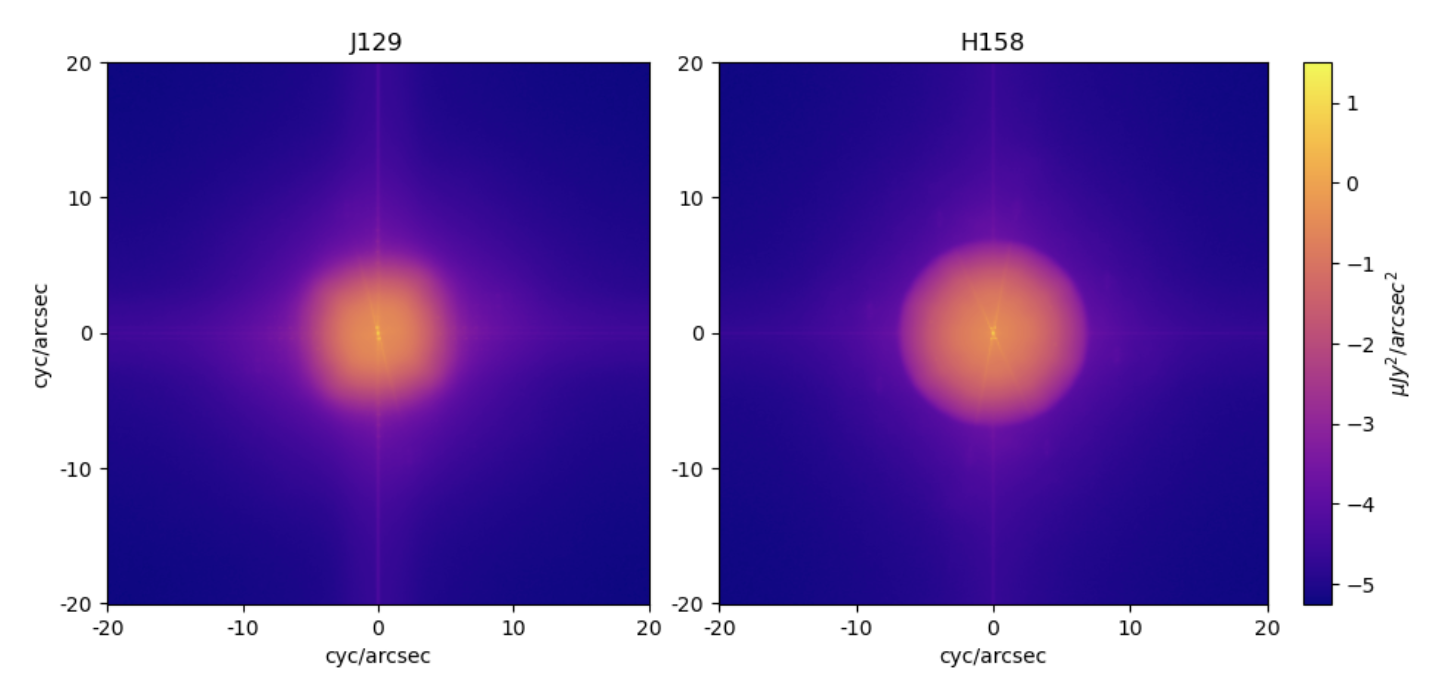}
    \caption{\label{fig:t23-2dspectra} The 2D averaged power spectrum of the Triplet lab noise fields in J and H bands, plotted on a logarithmic scale. The color scale shows the power $P(u,v)$ in units of $\mu$Jy$^2$ arcsec$^{-2}$. The minimum and maximum values of each power spectrum is as follows: {$3.4 \times 10^{-6}$} to {$3.1 \times 10^{1}$} for J129; and {$5.6 \times 10^{-6}$} to {$3.2 \times 10^{1}$} for H158.}
\end{figure*}

The spectra are shown in Figure \ref{fig:t23-2dspectra}.


As with the OU24 spectra, the bright central regions in these spectra come from limits of the output PSF specified. However, the T23-sims used a different output PSF: an airy disk convolved with a Gaussian (see H24 for details). In J129, we can see that the Gaussian dominate the output PSF as the central power smoothly fades into lower power in the high-wave number regions. In contrast, H158 is dominated by the airy disk, which determines a strict cutoff for the output PSF due to the detector range. This set cutoff causes the hard set circle rather than the smooth descent in power, as any signal outside the band limit is considered noise and set to zero by \Imcom\,. 



Most of the features in these spectra are the same as those seen above in the OU24 noise spectra, with the exception of two: the large + sign feature and the bright spots along the central Xs. 

The `+ sign' feature seen in both bands is a result of the repeated block boundaries. 
The boundaries create a feature similar to a step-function in real space which then results in the continuous lines seen in Fourier space. In both filters, the vertical line is slightly more distinct than the horizontal, and the `+ sign' is more distinct in J129. In the output images, there is a slight discontinuity in the stripes caused by noise from the detector. This discontinuity is likely greater at the horizontal boundaries than at the vertical boundaries, leading to a higher power in the vertical lines of the power spectrum. In the OU24 analysis, we implement a window function when making the blocks' power spectra to avoid this repeated step-function boundary. As can be seen in \S \ref{subsec:noise-2D}, this effectively removes the `+ sign' feature.

As in the model 1/f power spectra from Y24, the spectra all reach maximum power in the center. However, unlike the models, we see that there are several other features of higher power just outside the center, which lie along the X, seen clearly in Fig.~\ref{fig:t23-2dspectra}. The features are point-like in Fourier space, and thus are completely distributed as a sinusoidal signal in real space. The points also appear to be very small with little to no transverse smearing, which also tells us that most of the channels are likely picking up the same signal (the channel width is 14 arcsec, so corresponds to a width in Fourier space of 0.07 cycles/arcsec).

We estimate the features to be about 1 pixel wide, and 3 pixels from the center of the power spectrum. Due to the $8\times8$ binning, this would be 24 pixels in the detectors. With a theoretical dwell time per pixel of 5 $\mu$s, and 128 columns, we can calculate the time to read a row. Considering the entirety of the rows, including the overhead pixels, we have a block with a side length of 2600 pixels.
\begin{equation}
\frac{24 \cdot 0.11}{2600 \cdot 0.025 \cdot 128 \cdot 5 \mu \text{s}} = 6.35 \times 10^{-5} \text{MHz} = 63.5 \text{~Hz}.
\end{equation}
Using the actual read time (including overheads) results in a value of 57 Hz. We suspect that this feature is caused by the 60 Hz frequency of the North American power grid; this frequency and sometimes its harmonics have been observed in test data in the past. Fortunately, this source of interference will not be present in flight, with the spacecraft on direct current power from the solar arrays. Indeed, they are not even present in the FPT noise. Nevertheless, these differences are a reminder that it will be important to characterize the noise with in-flight dark exposures in order to inform the noise de-biasing pipeline.


\begin{figure*}[h]
    \centering
    \includegraphics[width=6.5in]{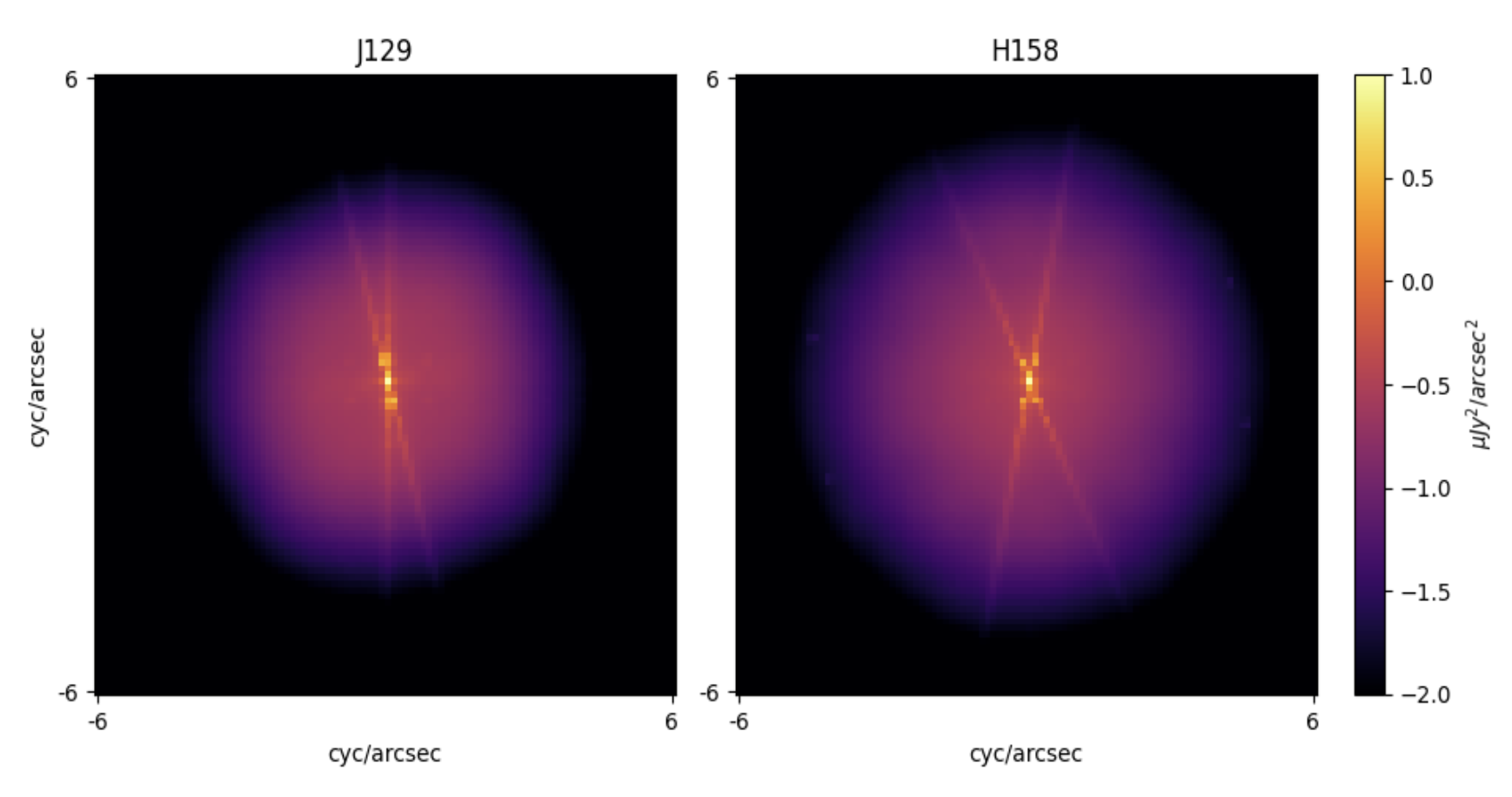}
    \caption{\label{fig:t23-zoomedspectra} Zoomed-in images of the central regions of the 2D Triplet noise power spectra. With the smaller scale for the power, the central features are more easily seen. Both bands have the highest power in a central spot of the power spectra.}
\end{figure*}

\subsection{1D Spectra}

The 1D spectra were also generated following the same methodology as used for the OU24 versions presented in \S \ref{subsec:noise-1D}. In this first version of the analysis, however, we performed an additional step to investigate whether the non-uniformity of sky coverage from the \textit{Roman} survey strategy would impact noise correlations. 

Uneven coverage of the sky results in uneven precision in measurements taken in different regions or on different scales. The power spectra represent the second moment of the images, and so any features we do (or do not) observe in these spectra will be affected by the coverage map in each block region.

\begin{figure*}[h]
    \centering
    \includegraphics[width=6.5in]{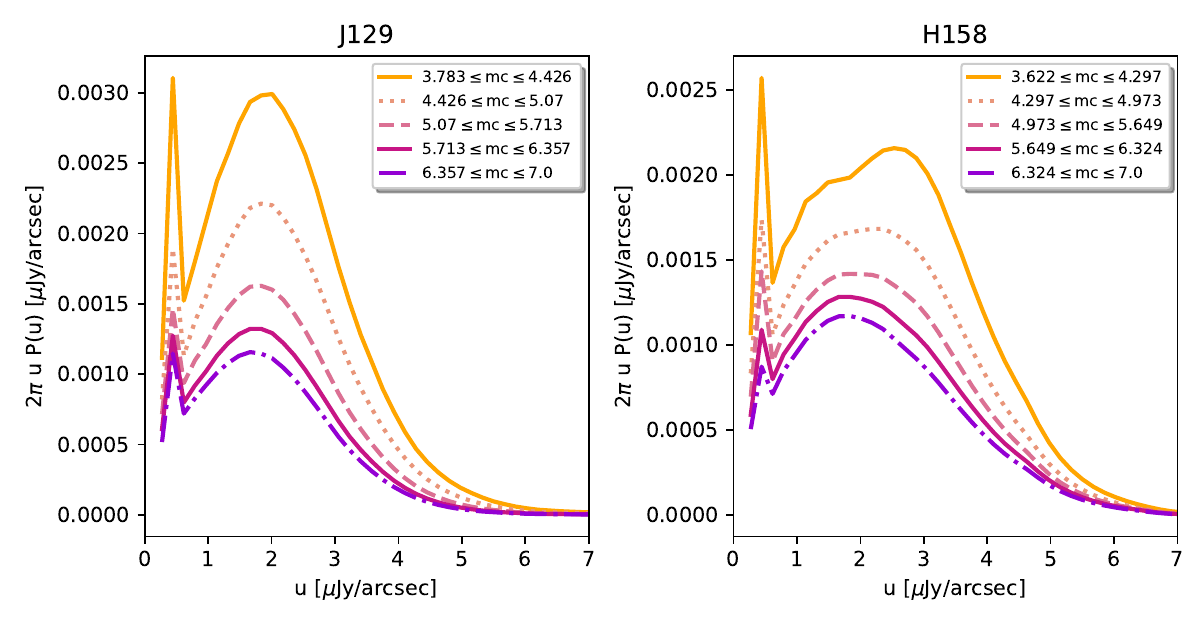}
    \caption{\label{fig:T23-1dspectra} The 1D azimuthally-averaged power spectrum of the lab noise fields in the J and H bands. Each line denotes a different bin of mean coverage.}
\end{figure*}

To understand the nature of these effects, we binned the 2D power spectra of each block in each observing filter by the average number of exposures that will be taken over a given region: a quantity we define as the ``mean coverage.'' The mean coverage range of each block was calculated and then subdivided into five even bins. Figure \ref{fig:T23-1dspectra} shows the mean coverage-binned, azimuthally averaged power spectra. 

The average behavior of the 1D spectra is the same as in the OU24 simulations, showing clearly an overall Gaussian shape with a small uptick for small wavenumbers caused by noise correlations on small scales. This figure additionally illustrates that regions with lower levels of coverage will have stronger noise correlations.

\subsection{Additive Bias Estimation}

\begin{figure*}[h]
    \centering
    \includegraphics[width=6.5in]{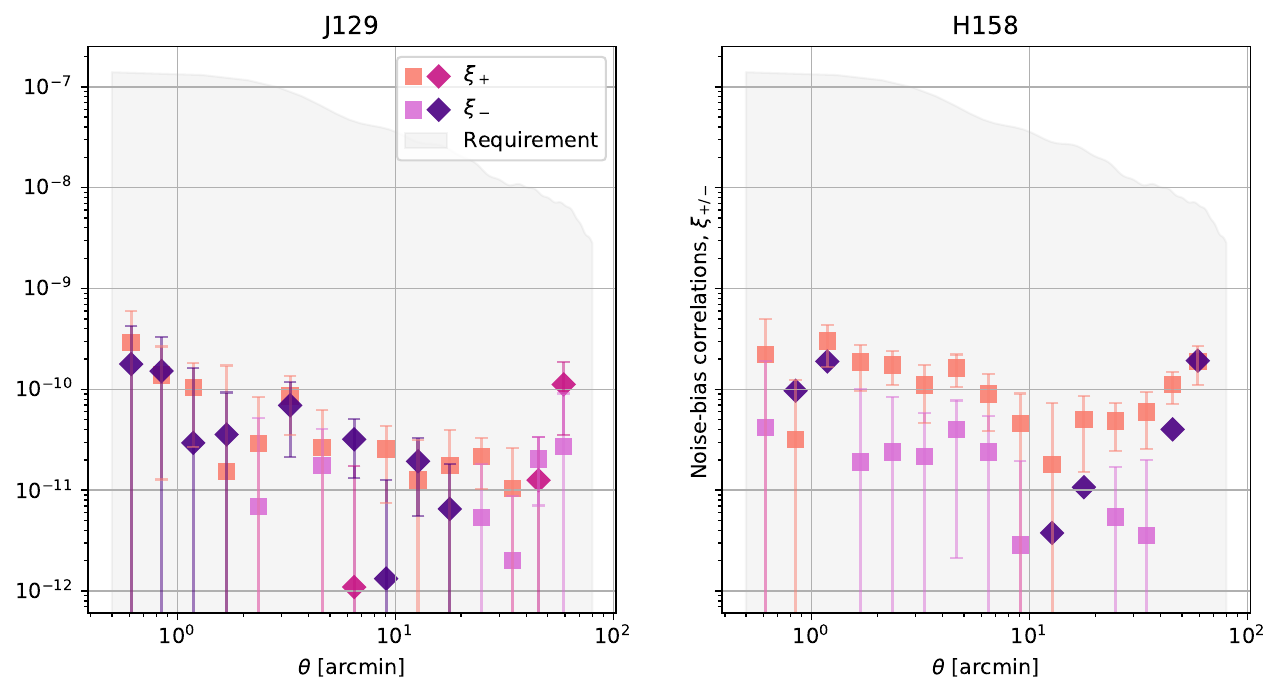}
    \caption{\label{fig:T23-ggcorrelations} Raw noise bias auto-correlations ($\Delta g$) estimated for the lab noise fields in J and H bands. Square markers indicate positive-valued correlations, while diamond markers indicate negative-values correlations, but plotted as the absolute value. }
\end{figure*}

Finally, we calculated the correlations in read noise biases on measured shapes of a grid of ``injected stars,'' which are described in detail in Y24. We normalized the stars and lab noise to magnitude $m_{\rm AB}=23$ as a starting magnitude for this analysis. Other than this difference, the noise bias correlations analysis follows the same format as described in \S \ref{sec:noise_corr}. Results are shown in Figure \ref{fig:T23-ggcorrelations}.

As with the OU24 noise bias analysis, we find that injected stars show very little noise biasing on shapes. Comparing to the same estimation of precision requirement shows that star shape noise biases are well below the requirement levels for \textit{Roman}.

\section{Derivation of Ellipticity Components}\label{app:e_derivation}

In our preliminary investigations of the choice and weighting of frames, we investigated the ellipticity bias that would be caused by a given noise field as a function of the position of the galaxy affected. For setting the optimal weighting of frames, we do this computation only for the simplest case, i.e., a round Gaussian object.

Using the notation of \cite{Bernstein2002}, we define our input image intensity as $I({\boldsymbol x})$. Taking some matrix of shears $\textbf{S}$ we can map the image to a sheared version,  $\tilde{I} = I(\textbf{S}{\boldsymbol x})$. We also define our measured moments ${M}(I)$, which will test for the amount of ellipticity in our image.
Using these quantities, we can define 
\begin{equation}
\delta M = \int d^{2}{\boldsymbol x} {W}({\boldsymbol x})(\Tilde{I}(x)-I(x)).
\label{eq:DM1}
\end{equation}
where ${W}$ is some set of 2D weight functions. Optimization of $\delta M$ will result in the most precise shape measurements.

Decomposing the image into multipole moments,
\begin{equation}
I(\textbf{x})=I(r,\theta)=\sum_{m=-\infty}^{\infty}I_{m}(r)e^{im\theta}
\end{equation}
with
\begin{equation}
    I_m(r)=\frac{1}{2\pi}\int_{0}^{2\pi}d\theta\; I(r,\theta)e^{-im\theta},
\end{equation}
and substituting into Eq.~(\ref{eq:DM1}) gives
\begin{equation}
    \delta M = \sum_{m=-\infty}^{\infty} \int_{0}^{\infty}r\,dr\; w_{m}(r)[\Tilde{I}_{m}(r)-I_{m}(r)],
\end{equation}
where the multipole weights $w_m(r)$ are an arbitrary radial function. Since object shape is determined by the quadrupole term, we will keep only the $m=2$ terms in this expansion. Using simplifications by assuming round isophotes in $I(x)$ and $w_{2}(r)=r^{2}w(r)$ (which can be seen by calculating the quadrupole term of ${M}(I)$) our equation simplifies to
\begin{equation}
    \delta M = -\frac{\eta}{4}\int_{0}^{\infty}dr\; r^{4}w(r)I'_{0}(r).
\end{equation}
For elliptical Gaussian objects, $w(r)=e^{-r^{2}/2\sigma^{2}}$
and $I_0(r) = (2\pi\sigma^2)^{-1} e^{-r^2/2\sigma^2}$,
so
\begin{equation}
    \delta M = \frac{\eta f}{8\pi\sigma^{4}}\int_{0}^{\infty}dr\; r^{5}e^{-r^{2}/\sigma^{2}}
= \frac{\eta f\sigma^{2}}{8\pi}.
\label{eq:15}
\end{equation}
This is of the form $\delta M = C\eta$ with $C=f\sigma^{2}/8\pi$.

Finally, the variance in shear components $\Delta\eta=\delta M/C$ can be solved for by integrating $\delta M$ assuming a Gaussian distribution on the measurement M. Converting coordinates back to Cartesian, we find
\begin{equation}
    e_{1}=\frac{4}{f\sigma^{2}}\int (\Delta x^{2}-\Delta y^{2}) e^{-\frac{\Delta x^{2}+\Delta y^{2}}{2\sigma^{2}}} I(x-\Delta x, y-\Delta y)\,d\Delta x\,d\Delta y
\label{eq:de1-res}
\end{equation}
and
\begin{equation}
    e_{2}=\frac{4}{f\sigma^{2}}\int (2\Delta x \Delta y) e^{-\frac{\Delta x^{2}+\Delta y^{2}}{2\sigma^{2}}} I(x-\Delta x, y-\Delta y)\,d\Delta x\,d\Delta y.
\label{eq:de2-res}
\end{equation}.

These equations for the ellipticity components are used in section \ref{sec:labdata} to estimate shear variance from noise and select which frames to keep from the lab noise dark exposures.

\label{lastpage}

\end{document}